\documentclass[conference]{IEEEtran}
\IEEEoverridecommandlockouts
\usepackage{cite}
\usepackage[T1]{fontenc}
\usepackage[utf8]{inputenc}
\usepackage{amsmath,amssymb,amsfonts}
\usepackage{amsthm}
\usepackage{algorithm}
\usepackage{algorithmicx}
\usepackage{algpseudocode}
\usepackage{graphicx}
\usepackage{textcomp}
\usepackage{xcolor}
\usepackage{booktabs}
\usepackage{multirow}
\usepackage{hyperref}
\usepackage{url}
\usepackage{orcidlink}
\usepackage{hyperref}

\providecommand{\botrule}{\bottomrule}

\theoremstyle{plain}
\newtheorem{theorem}{Theorem}
\newtheorem{proposition}[theorem]{Proposition}
\newtheorem{corollary}[theorem]{Corollary}
\newtheorem{lemma}[theorem]{Lemma}

\theoremstyle{definition}

\newtheorem{remark}{Remark}

\begin{document}

\title{MARGIN: Margin-Aware Regularized Geometry for Imbalanced Vulnerability Detection}

\author{
\IEEEauthorblockN{
Yuteng Zhang\orcidlink{0009-0006-4931-8003},
Huifang Ma\orcidlink{0000-0002-5104-8982},
Jiahui Wei\orcidlink{0000-0001-6762-7180},
Qingqing Li\orcidlink{0000-0001-8114-5831},
Yafei Yang\orcidlink{0009-0008-3679-2720},
Xin Ji\orcidlink{0009-0001-8706-4300}
}
}

\maketitle

\begin{abstract}
Software vulnerability detection is critical for ensuring software security and reliability. Despite recent advances in deep learning, real-world vulnerability datasets suffer from two severe challenges: frequency imbalance and difficulty imbalance. We reinterpret these challenges from an embedding geometry perspective, observing that such imbalances cause geometric distortions in the hyperspherical representation space. To address this issue, we propose \textbf{MARGIN} (\textbf{M}argin-\textbf{A}ware \textbf{R}egularized \textbf{G}eometry for \textbf{I}mbalanced Vulnerability Detectio\textbf{N}), a metric-based framework that learns discriminative vulnerability representations through adaptive-margin metric learning and hyperspherical prototype modeling. MARGIN adjusts geometric regularization according to the distribution structure estimated by the von Mises--Fisher concentration. It contracts each class's probability mass region, i.e.\ its Learned Voronoi Cone, toward the Ideal Voronoi Cone of the balanced partition. This reduces geometric distortion and yields more stable decision boundaries. Extensive experiments on public vulnerability datasets show that MARGIN consistently outperforms strong baselines, achieving notable improvements in classification and detection, especially on challenging, imbalanced datasets. Further analysis shows that MARGIN produces more structured embedding geometries, improving robustness, interpretability, and generalization.
\end{abstract}


\section{Introduction}

As modern software systems keep growing in size and complexity, security vulnerabilities hidden in source code have become a serious threat to software reliability and system safety. These vulnerabilities can cause unexpected program failures or be exploited by attackers to break into systems. Therefore, automatically detecting and classifying vulnerabilities from source code has become an important problem in software engineering.

Recent advances in deep learning have significantly improved software vulnerability detection and classification. Despite these advances, such models still face substantial challenges when applied to real-world vulnerability datasets. Deep learning-based vulnerability analysis aims to learn discriminative class representations by clustering samples of the same category while separating different categories. In practice, real-world vulnerability data are often highly imbalanced in both class frequency and sample difficulty, which can bias model optimization, hinder reliable convergence, and degrade generalization performance.

\textbf{Frequency Imbalance.} A key challenge in deep learning for vulnerability detection is the severe \textit{frequency imbalance} in real-world software data. This issue appears not only as the imbalance between vulnerable and non-vulnerable code, but more importantly as the uneven distribution across different \textit{Common Weakness Enumeration (CWE)} categories.

\begin{figure*}[t]
	\centering
	\includegraphics[width=\textwidth]{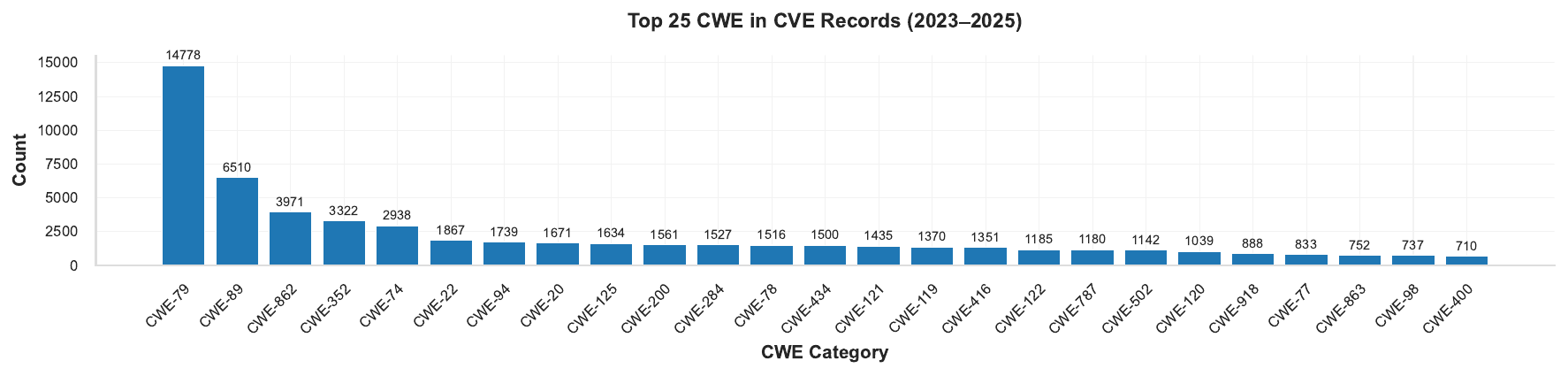}
	\caption{\textbf{Top 25 CWE Types in CVE Project (2023--2025).} Frequency distribution of CWE categories in publicly disclosed vulnerabilities from the CVE Project. The distribution is highly imbalanced, where a small number of CWE types account for a large portion of vulnerabilities while many others appear only rarely. This long-tailed pattern reflects the uneven real-world occurrence of software weaknesses and highlights the class imbalance in vulnerability datasets.}
	\label{fig:cwe-statistics}
\end{figure*}

As shown in Figure~\ref{fig:cwe-statistics}, which presents the frequencies of the top 25 CWE types from recent Common Vulnerabilities and Exposures (CVE) records (2023--2025)~\cite{MITRE_CWE_Top25_2025}, the data follow a clear long-tailed distribution with large differences in sample counts. A small number of common categories, such as CWE-79 (Cross-Site Scripting) and CWE-89 (SQL Injection), contain tens of thousands of samples and make up a large portion of the dataset. In contrast, many other vulnerabilities, such as CWE-400 (Uncontrolled Resource Consumption), CWE-98 (Improper Control of Filename for Include/Require Statement), and CWE-863 (Incorrect Authorization), appear very rarely and only account for a small part of the data.

\begin{figure}\centering \includegraphics[width=0.6\columnwidth]{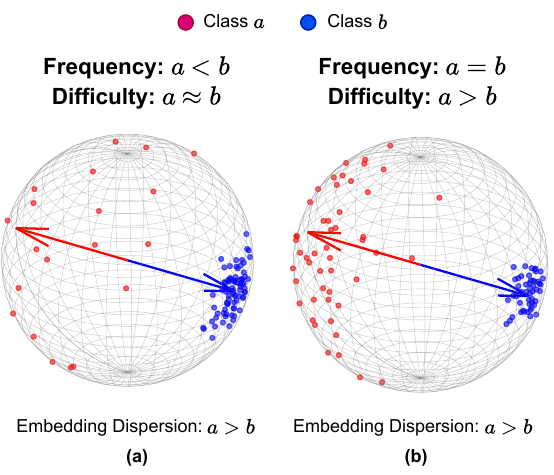} \caption{
\textbf{(a)} With comparable class difficulty, lower-frequency classes yield embeddings that converge less reliably around their prototypes on the manifold.
\textbf{(b)} With equal class frequencies, higher difficulty (i.e., greater intra-class variance) results in embeddings that converge less reliably around their prototypes on the manifold.
}
\label{fig:imbalance-effect}
\end{figure}

This large gap in frequency leads to biased metric learning. Standard loss functions such as cross-entropy and contrastive loss treat all samples equally~\cite{surveyofmetriclearningloss}, so classes with more samples contribute much more to the gradients and dominate the parameter updates. Meanwhile, minority classes provide much weaker signals and have limited influence on the learned representations. In the learned embedding space, majority-class samples typically cluster tightly, while minority-class samples remain scattered and fail to reliably converge around their prototypes. As illustrated in Figure~\ref{fig:imbalance-effect}(a), even when class difficulty is similar, the embeddings of the minority class $a$ show a more dispersed and blurred distribution than class $b$.

Because of this, the model tends to learn biased patterns from high-frequency classes, shifting the decision boundary toward dominant categories while overlooking the subtle characteristics of rare vulnerabilities~\cite{AnEmpiricalStudyoftheImbalanceIssueinSoftwareVulnerabilityDetection,ncforimbalance}. Such bias severely limits the reliability of vulnerability detection systems in real-world security scenarios, where accurately identifying rare yet critical vulnerabilities is particularly important.

\begin{figure*}[t]
	\centering
	\includegraphics[width=\textwidth]{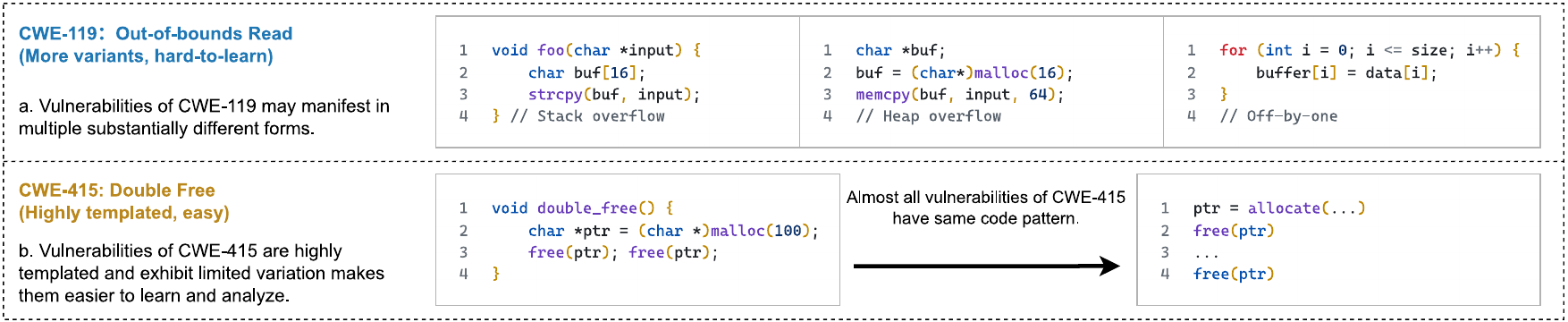}
	\caption{\textbf{Illustration of vulnerability diversity and learning difficulty across two CWE types.}
(a) CWE-119 shows high embedding variance with diverse patterns, including stack overflows, heap overflows, and off-by-one errors.
(b) In contrast, CWE-415 follows a highly templated \texttt{allocate$\rightarrow$free$\rightarrow$free} pattern with limited variation.}
	\label{fig:learning-difficulty}
\end{figure*}

\textbf{Difficulty Imbalance.} Most existing methods for addressing imbalanced vulnerability detection primarily focus on frequency imbalance. Common solutions include resampling, cost-sensitive learning, Focal Loss, and class-balanced reweighting~\cite{surveyofmetriclearningloss,Cui_2019}. While these approaches reduce the dominance of majority classes, they generally treat samples from different classes as having comparable learning difficulty. In practice, however, different CWE categories can show very different degrees of implementation heterogeneity, which can lead to different learning difficulties even when their class frequencies are balanced.

The underlying source of this difficulty imbalance lies in the diversity of vulnerability manifestations within different CWE categories. Although CWE provides a standardized taxonomy, vulnerabilities belonging to the same CWE can arise from very different syntax, control-flow structures, data-flow dependencies, and usage contexts~\cite{code2vec}. Consequently, different CWE categories may show markedly different distributions of implementation patterns. For example, as shown in Figure~\ref{fig:learning-difficulty}, CWE-119 (Improper Restriction of Operations within the Bounds of a Memory Buffer) can manifest as stack overflows, heap overflows, or off-by-one loop errors, resulting in substantial structural diversity. In contrast, CWE-415 (Double Free) often shows more repetitive patterns, such as repeatedly freeing the same pointer. Such differences in structural and semantic heterogeneity can be described by code-level statistics~\cite{investigation_for_datasets}.

This implementation heterogeneity is subsequently reflected in the feature space produced by the backbone encoder. Given a fixed encoder, samples with more diverse structural and semantic patterns are likely to be mapped to a more dispersed region of the representation space, whereas samples showing more consistent patterns tend to produce more compact representations. We therefore use intra-class feature variance to describe the representation-level manifestation of class difficulty; the disparity in variance scale across classes drives difficulty imbalance. If all classes showed comparably large (or comparably small) dispersion, a shared optimization pressure could in principle contract them at similar rates. Intra-class variance differs greatly across CWE categories, with some classes markedly dispersed and others tightly clustered. It is in this case that cross-class heteroscedasticity translates into asymmetric optimization dynamics.

Differences in variance scale across classes further affect the optimization dynamics. Classes with comparatively compact and consistent representations can be organized around their class prototypes more readily, whereas classes with comparatively larger intra-class dispersion require the model to reconcile more heterogeneous representations and may therefore show slower or less stable convergence~\cite{Geman_1992}. Under comparable class frequencies, this scale disparity leads to different rates at which individual classes contract toward their respective prototypes. Thus, implementation heterogeneity is propagated through the learned representations into class-dependent convergence behavior only insofar as it is unevenly distributed across classes.

The resulting convergence differences ultimately produce different geometric structures in the embedding space. Easier classes tend to form compact and well-separated clusters, while more difficult classes can retain larger intra-class dispersion, weaker prototype concentration, or greater overlap with neighboring classes. This phenomenon is also consistent with the observation of neural collapse, where well-optimized classes tend to form increasingly compact clusters in the terminal feature space~\cite{papyan2020prevalence}.
Accordingly, difficulty imbalance can be viewed as a geometric consequence of heterogeneous learning dynamics, and not merely as a property of the class labels. As illustrated in Figure~\ref{fig:imbalance-effect}(b), even when class frequencies are identical, the difficult class $a$ can show a more dispersed feature distribution than class $b$. As training proceeds, the two classes may also contract toward their respective prototypes at different rates, resulting in distinct geometric configurations. When frequency imbalance and difficulty imbalance coexist, these differences in class-wise geometry become coupled with unequal sample frequencies. Therefore, the imbalance problem is reflected not only in the number of samples assigned to each class, but also in how those samples are distributed around their class prototypes and how this distribution evolves during optimization.

\textbf{Lack of Interpretability.} Most existing approaches mainly aim to improve classification performance, while paying less attention to interpretability. Common evaluation metrics such as the Matthews Correlation Coefficient (MCC) and the F1-score reflect only prediction results and cannot show whether the embedding space has clear and well-discriminated geometric structures under severe imbalance~\cite{AnEmpiricalStudyoftheImbalanceIssueinSoftwareVulnerabilityDetection}. Without explicit constraints on representation geometry, models may depend on locally optimal but unstable decision boundaries, which can be highly sensitive to distribution shifts or rare vulnerability patterns. For security-critical tasks such as vulnerability detection, such instability can greatly reduce practical reliability.

In essence, both frequency imbalance and difficulty imbalance can be viewed as one shared geometric issue in the embedding space, namely the uneven concentration of class distributions. This work shows that imbalance in vulnerability detection is closely related to distortion in the learned embedding geometry~\cite{ncforimbalance,Cui_2019}. Frequency and difficulty imbalance cause dominant classes to show highly dispersed and distorted distributions in the learned decision space. Therefore, explicitly modeling the embedding space in metric learning from a geometric and manifold perspective offers a more principled way to improve both robustness and interpretability.

To address these challenges, we propose \textbf{MARGIN} (\textbf{M}argin-\textbf{A}ware \textbf{R}egularized \textbf{G}eometry for \textbf{I}mbalanced Vulnerability Detectio\textbf{N}), a metric-based framework that revisits vulnerability detection from a hyperspherical embedding perspective. It provides a unified geometric view in which both frequency and difficulty imbalance are described by how concentrated each class is on the hypersphere. By explicitly shaping the geometry of the embedding space, MARGIN reduces representation bias through adaptive margin learning and prototype-based regularization. This helps stabilize decision boundaries and improve the separation between classes under severe imbalance. Overall, the framework connects practical imbalance handling with a clear geometric interpretation, leading to better performance and improved interpretability.

The contributions of this paper are summarized as follows:

\begin{enumerate}
\item We revisit two critical imbalances in vulnerability detection, i.e., \textit{frequency imbalance} and \textit{difficulty imbalance}, and analyze their effects in an interpretable manner from the perspective of embedding geometry. (Section~\ref{sec:motivation-analysis})
\item We propose \textbf{MARGIN}, a metric-learning framework that integrates adaptive mechanisms and hyperspherical prototype modeling to reduce both imbalances simultaneously. (Section~\ref{sec:methodology})
\item Extensive experiments on multiple public vulnerability datasets show that MARGIN consistently improves vulnerability detection and classification performance. (Sections~\ref{sec:experiments} and \ref{sec:case-study})
\end{enumerate}

\section{Related Work}

\subsection{Static Vulnerability Analysis}

Before the emergence of deep learning-based approaches, software vulnerability detection primarily relied on static program analysis with manually specified rules and expert knowledge. Representative techniques include rule-based analysis, pattern matching, and structural analyses over ASTs and control-flow graphs (CFGs)~\cite{babati2017static}. Widely used tools such as CppCheck, CodeQL, Semgrep, and Clang-Tidy apply these techniques to identify predefined vulnerability patterns and program properties. Such approaches offer strong interpretability and have been widely adopted in industrial practice. However, their effectiveness depends heavily on the completeness and quality of manually designed rules, making them difficult to adapt to evolving vulnerability patterns and diverse large-scale software systems. More sophisticated analyses, such as symbolic execution and path-sensitive analysis, can provide deeper reasoning but often face substantial scalability challenges, including path explosion. Meanwhile, rule-based approaches remain constrained by incomplete vulnerability specifications and the cost of continuously maintaining analysis rules~\cite{Context-flow-andfield-sensitive}. These limitations have motivated the development of learning-based vulnerability detection methods, from early systems such as VulDeePecker~\cite{vuldeepecker} to recent deep models~\cite{LIANG2025104098}, which learn discriminative patterns directly from labeled code and reduce reliance on manually crafted vulnerability rules.

\subsection{Representation Learning}

Code representation learning maps programs into continuous embedding spaces for downstream tasks such as vulnerability detection and code search. Early approaches, such as Code2Vec~\cite{code2vec} and weighted word-vector models~\cite{TCNN_BiGRU}, rely on token- or path-based representations and largely overlook global structural dependencies. Graph-based methods, including MGVD~\cite{mgvd} and Devign~\cite{devign}, incorporate structural information from abstract syntax trees, control-flow graphs (CFGs), and data-flow graphs, improving semantic expressiveness at the cost of higher computational complexity. More recently, Transformer-based pretrained models, such as CodeBERT~\cite{codebert}, GraphCodeBERT~\cite{graphcodebert}, CodeT5~\cite{codet5}, UniXCoder~\cite{unixcoder}, and VulBERTa~\cite{vulberta}, have become the dominant approach due to their strong transferability. Despite their effectiveness, existing studies primarily evaluate learned representations through aggregate performance metrics, paying limited attention to embedding-space geometry and class-wise distributional properties~\cite{lian2026ethical}. This gap is particularly relevant given recent findings that aligned language models remain vulnerable at the semantic representation level, with such weaknesses transferring across models~\cite{lian2025semantic,lian2026ethical}, alongside systematic robustness benchmarks for deep models in other application domains~\cite{mei2024robustness}. Because MARGIN builds upon pretrained code representations instead of learning code features from scratch, it inherits the semantic structure of the encoder's embedding space and operates on its $\ell_2$-normalized form. Accordingly, the geometric and distributional properties of this space directly affect the resulting vulnerability representations. MARGIN addresses this issue by incorporating representation-level geometric analysis into the training objective, explicitly penalizing the excess of the Learned Voronoi Cone over the Ideal one to reduce uneven angular concentration under class imbalance.

\subsection{Imbalanced Classification Learning}

Class imbalance remains a major challenge in vulnerability detection. Conventional methods, such as Weighted Cross-Entropy, Focal Loss~\cite{focalloss}, LDAM~\cite{ldam}, and Class-Balanced Loss~\cite{AnEmpiricalStudyoftheImbalanceIssueinSoftwareVulnerabilityDetection,Cui_2019,surveyofmetriclearningloss}, reduce imbalance through reweighting strategies, while data-level methods such as fuzzy oversampling~\cite{deepbalance} rebalance the training distribution instead. Both families generally assume homogeneous intra-class distributions and lack explicit geometric constraints~\cite{He2024EnhancingDL,DoesDataSamplingImproveDeepLearning-BasedVulnerabilityDetection,AnEmpiricalStudyoftheImbalanceIssueinSoftwareVulnerabilityDetection,Cui_2019}. Recent work explores prototype learning and metric-space regularization to improve minority discrimination. However, such geometry-aware approaches remain underexplored in software vulnerability detection. Existing studies typically treat representation learning, metric learning, and imbalance handling as separate problems. In contrast, multi-CWE vulnerability detection requires jointly addressing semantic heterogeneity, long-tailed distributions, and embedding geometry. This work unifies these aspects by integrating hyperspherical modeling, adaptive margins, and prototype-based learning into a single framework.

\section{Preliminaries}

\subsection{Problem Formulation}

Consider a vulnerability dataset
$\mathcal{D} = \{(x_i, y_i)\}_{i=1}^{N}$,
where each sample consists of a code snippet $x_i \in \mathcal{X}$ and a class label
$y_i \in \mathcal{Y}$. The label space is defined as
$\mathcal{Y} = \{\texttt{Non-Vul}, \texttt{CWE-a}, \texttt{CWE-b}, \dots\}$,
where \texttt{Non-Vul} denotes non-vulnerable code and each \texttt{CWE-*} represents a specific vulnerability category defined in the CWE taxonomy.

This task is formulated as a \textit{single-label multi-class classification} problem in a closed-set setting, where each code sample is associated with exactly one label. In practice, real-world datasets show severe frequency and difficulty imbalance across different labels.

\subsection{Cosine Softmax Loss}

To address class imbalance in vulnerability detection from a geometric perspective, we adopt a unit hyperspherical prototype-based classification framework. This formulation defines a structured manifold that can be explicitly analyzed and regularized, allowing imbalance to be tackled in a unified geometric manner. Each class is represented by a prototype, serving as a stable and interpretable semantic center in the embedding space~\cite{Ranjan2017L2constrainedSL,sphericaltextembedding}.

We begin with the standard \emph{Cosine Softmax Loss} for single-label multi-class classification. For each input, the encoder produces a $d$-dimensional feature, and we apply $\ell_2$-normalization to both the feature $\mathbf{e} \in \mathbb{R}^d$ and the classifier weight $\mathbf{w}_i \in \mathbb{R}^d$. The linear classifier then produces logits for each class, followed by the standard softmax cross-entropy loss:

\begin{equation}
z_i = \mathbf{\tilde{w}}_i^\top \mathbf{\tilde{e}},
\quad
\ell_{i}
=
- \log
\frac{\exp(z_i)}
{\sum_{j=1}^{|\mathcal{Y}|} \exp(z_j)},
\quad i \in \{1,\dots,|\mathcal{Y}|\}
\end{equation}

The logit $z_i$ can also be written as:

\begin{equation}
z_i=\cos \theta_i
\end{equation}
where $\theta_i$ denotes the angle between $\mathbf{\tilde{w}}_i$ and $\mathbf{\tilde{e}}$. Thus, the decision score depends purely on angular similarity, and vector magnitude variations are eliminated.

We introduce the scaling factor $s_0$ to sharpen the logits~\cite{Kobayashi_2025_ICCV}, and the resulting Cosine Softmax Loss is defined as:
\begin{equation}
\ell_i =
-
\log
\frac{\exp(s_0 \cos \theta_{i})}
{\sum_{j=1}^{|\mathcal{Y}|}
\exp(s_0 \cos \theta_{j})}
\label{form:normalized-cos-softmax-loss}
\end{equation}

The Cosine Softmax Loss makes both features and class weights lie on the unit hypersphere
$\mathbb{S}^{d-1}$. The vector $\tilde{\mathbf{w}}_i$ is interpreted as the \emph{weight prototype} of class $i$.
Training therefore encourages samples to align with their corresponding prototypes, forming compact clusters separated by angular margins. This geometric view enables explicit analysis of imbalance effects in the embedding space.

\subsection{Voronoi Cell and Neural Collapse}

\begin{figure}\centering \includegraphics[width=0.6\columnwidth]{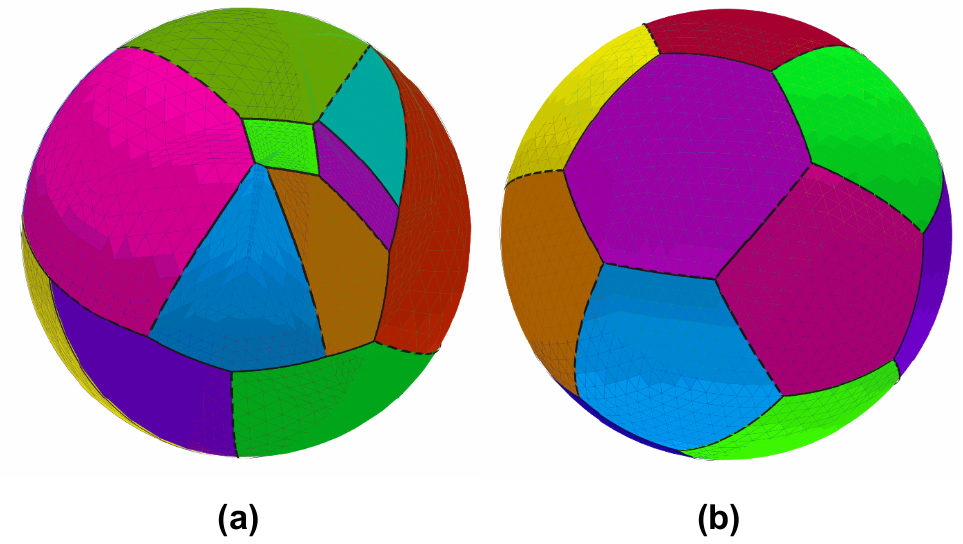} \caption{
\textbf{(a)} Under joint frequency and difficulty imbalance, the prototypes deviate from the symmetric configuration, so the learned Voronoi cells they induce become uneven in size; the probability mass of a class then spills outside its own decision boundary into those of its neighbors.
\textbf{(b)} With \emph{Ideal Voronoi Cells}, the partition is uniform and symmetric on the hyperspherical manifold and every class's probability mass stays inside its own cell, giving strong class separability.
}
\label{fig:voronoi-cells}
\end{figure}

Recent studies show that deep classifiers trained without bias terms on ideally balanced data tend to show the Equiangular Tight Frame (ETF) structure under the \emph{Neural Collapse} phenomenon during the later stage of training~\cite{markou2024guiding}. All weight prototypes converge to a highly symmetric geometric configuration, in which the pairwise similarities converge to a constant. The prototypes are then uniformly distributed on the unit hypersphere:

\begin{equation}
\tilde{\mathbf{w}}_i^\top \tilde{\mathbf{w}}_j
=
-\frac{1}{|\mathcal{Y}|-1},
\quad
\forall i\neq j
\label{form:etf}
\end{equation}

In this setting, the unit hypersphere $\mathbb{S}^{d-1}$ is partitioned into $|\mathcal{Y}|$ Voronoi cells by assigning every point to its closest weight prototype. When the prototypes satisfy Eq.~(\ref{form:etf}), all cells are congruent and subtend the same angle, as illustrated in Figure~\ref{fig:voronoi-cells}(b); these are called the \emph{Ideal Voronoi Cells}. Under joint frequency and difficulty imbalance, however, the learned prototypes deviate from this symmetric configuration, so the cells they induce become uneven and vary in width, leading to poor class separability, as illustrated in Figure~\ref{fig:voronoi-cells}(a).

When the number of classes $|\mathcal{Y}|$ is sufficiently large, each Voronoi cell is well approximated by a spherical cap on the hypersphere~\cite{Coxeter1969}. Taking a Voronoi cell as the base and the center of the unit hypersphere as the apex yields a \emph{spherical cone}. Throughout this paper, a spherical cone is described by its \emph{apex angle}. The cone whose base is an Ideal Voronoi Cell is the \emph{Ideal Voronoi Cone} (IVC). Its apex angle equals that of its base cell. It depends only on the number of classes and is given by

\begin{equation}
\theta^{\mathrm{ideal}}
=
\arccos\left(
-\frac1{|\mathcal{Y}|-1}
\right)
\label{form:ideal-angle}
\end{equation}

Because the $|\mathcal{Y}|$ Voronoi cells tile the hypersphere exactly, $\theta^{\mathrm{ideal}}$ fixes the angular budget available to each class: a cone centered on a prototype cannot grow beyond this apex angle without entering the territory of its neighbors. In the ETF configuration of Eq.~(\ref{form:etf}), every prototype occupies exactly the budget it is granted; otherwise the prototypes are pushed off center, and the crowded classes are confined to cells much narrower than the room their IVCs would grant them.

At the same time, the probability mass regions of embeddings for each class converge toward equal size and lie predominantly within the corresponding IVC defined by its weight prototype. Also, the mean feature prototype of each class aligns with its corresponding weight prototype. Let $\tilde{\boldsymbol{\mu}}_i$ denote the normalized mean feature of class $i$. Under Neural Collapse, $\tilde{\boldsymbol{\mu}}_i = \tilde{\mathbf{w}}_i$ for all $i \in \{1,\dots,|\mathcal{Y}|\}$.

Ideally, the ETF structure yields clearly discriminative and unbiased decision boundaries~\cite{ncforimbalance}. Under class-imbalanced conditions, however, such an ETF structure cannot be fully established. The prototypes remain angularly separated but lose their symmetry. As a result, the cells of the crowded classes are compressed well below the room their IVCs would grant them, while the embeddings of the hard classes spread beyond the room they are granted. This mismatch gives rise to the distortion region. This observation suggests that, to reduce the prediction bias induced by class imbalance from a geometric perspective, it is essential to explicitly encourage the embedding distribution to evolve toward an ETF structure.

\subsection{vMF Distribution for Hyperspherical Modeling}

Following the principle of maximum entropy, feature embeddings are represented on the unit hypersphere, and the normalized code representations are modeled with the von Mises--Fisher (vMF) distribution. This is a modeling prior and an inductive bias. Similar to the Gaussian assumption in Euclidean representation modeling, the vMF assumption provides a tractable description of directional representation distributions and enables explicit measurement of class-wise concentration. Accordingly, the problem is formulated as maximum likelihood estimation under the vMF distribution.

Among all distributions on $\mathbb{S}^{d-1}$ with a prescribed mean direction, the vMF distribution is the maximum-entropy choice~\cite{mardia2000directional}. In this framework, the concentration parameter $\kappa_i$ measures how tightly samples cluster around their mean direction $\boldsymbol{\mu}_i$~\cite{vmf,li2024probabilisticcontrastivelearningexplicit}. Formally, a normalized embedding $\mathbf{\tilde{e}} \in \mathbb{S}^{d-1}$ follows a vMF distribution $\mathbf{\tilde{e}} \sim \mathrm{vMF}(\boldsymbol{\tilde{\mu}}_i,\kappa_i)$ if

\begin{equation}
\label{form:vmf-distribution}
p(\mathbf{\tilde{e}} \mid \boldsymbol{\tilde{\mu}}_i,\kappa_i)
=
C_d(\kappa_i)
\exp\left(
\kappa_i \boldsymbol{\tilde{\mu}}_i^\top \mathbf{\tilde{e}}
\right)
\end{equation}

\noindent where

\begin{equation}
C_d(\kappa_i)
=
\frac{\kappa_i^{\frac d2 -1}}
{(2\pi)^{\frac d2}
I_{\frac d2 -1}(\kappa_i)}
\end{equation}

\noindent and $I_{\nu}(\cdot)$ denotes the modified Bessel function of the first kind.

A larger $\kappa_i$ indicates stronger concentration of embeddings around $\boldsymbol{\tilde{\mu}}_i$, leading to more compact class distributions.

Under the vMF assumption, the high-density region of each class on the hypersphere can be approximated as a spherical cap centered at $\boldsymbol{\tilde{\mu}}_i$. Taking this region as the base and the center of the unit hypersphere as the apex yields a spherical cone, referred to as the \emph{Learned Voronoi Cone} (LVC). The two cones now stand in contrast to each other: the Ideal Voronoi Cone of Eq.~(\ref{form:ideal-angle}) gives the angular extent a class is entitled to under a balanced partition, whereas the Learned Voronoi Cone gives the extent that the class actually occupies.

For a given confidence level $\alpha$, the corresponding high-probability region satisfies
\begin{equation}
P\left(
\arccos(\boldsymbol{\tilde{\mu}}_i^\top \mathbf{\tilde{e}})
\le \theta_i^{\mathrm{learned}}
\right)
=\alpha
\end{equation}

\noindent where $\theta_i^{\mathrm{learned}}$ denotes the apex angle of the Learned Voronoi Cone.

Importantly, for a fixed confidence level $\alpha$, the apex angle $\theta_i^{\mathrm{learned}}$ is solely determined by the concentration parameter $\kappa_i$. A larger $\kappa_i$ leads to a smaller angle, indicating tighter concentration around the prototype, whereas a smaller $\kappa_i$ produces a wider angle, reflecting more dispersed intra-class distributions.

\section{Motivation Analysis}
\label{sec:motivation-analysis}

\begin{figure*}\centering \includegraphics[width=\textwidth]{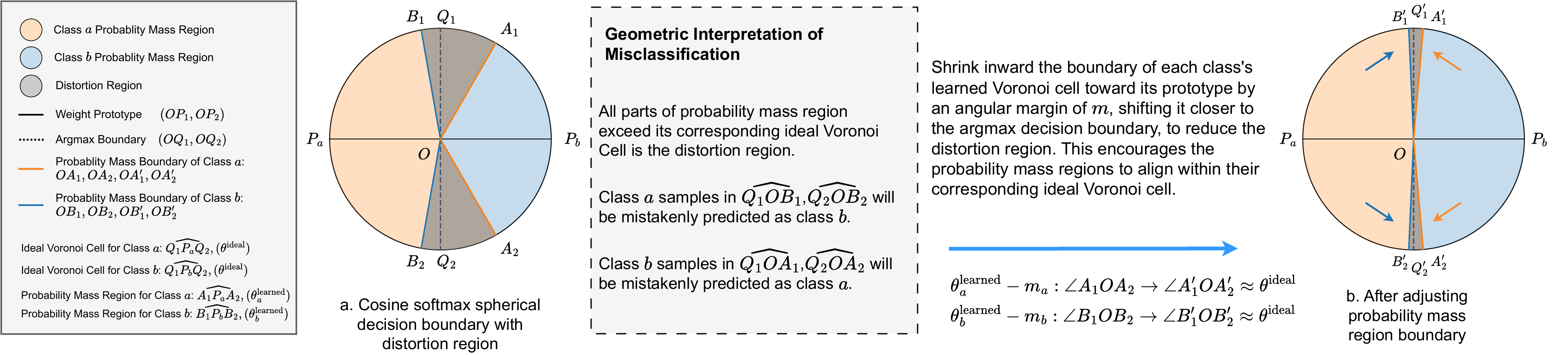} \caption{\textbf{Geometric Interpretation of Misclassification and Heuristic Design of MARGIN.} Conceptual illustration of the spherical angular spans underlying class decisions for the Cosine Softmax Loss and MARGIN on $\mathbb{S}^{1}$.}
\label{fig:problem-analysis}
\end{figure*}

We next explain why an uneven embedding distribution leads to poor class separability and how the proposed method aligns the probability mass region of each class with its Voronoi cell. For an intuitive illustration, consider the unit circle $\mathbb{S}^{1}$ as an example.

Under the dual imbalance of class frequency and difficulty, neural networks in the late training stage learn hyperspherical embeddings in which class prototypes are well separated by large angular distances. This stage corresponds to the Neural Collapse regime. As illustrated in Figure~\ref{fig:problem-analysis}, let $P_a$ and $P_b$ denote the prototypes of classes $a$ and $b$, respectively.

According to the argmax decision rule, the angular bisectors of $\angle P_aOP_b$ and $\angle P_bOP_a$, denoted as $OQ_1$ and $OQ_2$, partition the unit circle into the Voronoi cells of classes $a$ and $b$, namely $\widehat{Q_1P_aQ_2}$ and $\widehat{Q_1P_bQ_2}$. These are exactly the decision regions of the classifier, and they agree with the bases of the IVCs only when the prototypes attain the symmetric configuration of Eq.~(\ref{form:etf}).

However, under imbalance, class-conditional embeddings do not always concentrate tightly around their corresponding prototypes. Instead, they may deviate and extend beyond their associated cells, leading to misclassification~\cite{ncforimbalance}.

To describe this phenomenon, consider the binary classification scenario shown in Figure~\ref{fig:problem-analysis}. The embeddings of classes $a$ and $b$ can be reasonably assumed to follow vMF distributions on $\mathbb{S}^1$. Their probability mass regions can be viewed as spherical caps, namely $\widehat{A_1OA_2}$ and $\widehat{B_1OB_2}$, whose apex angles are $\theta^\text{learned}_a$ and $\theta^\text{learned}_b$, with $\theta^\text{learned}_a > \theta^\text{learned}_b$. This indicates that class $a$ shows a more dispersed embedding distribution.

Importantly, the Learned Voronoi Cone of a class may extend beyond the Ideal Voronoi Cone, i.e.\ beyond the angular budget that the symmetric partition grants it. Because both angles are measured in the same apex-angle convention, this happens precisely when $\theta^\text{learned}_i>\theta^\text{ideal}$. For class $a$, the exceeding portions $\widehat{Q_1OA_1}$ and $\widehat{Q_2OA_2}$ intrude into the territory of class $b$, namely into the cell $\widehat{Q_1P_bQ_2}$, leading to false negatives for class $a$ and false positives for class $b$. A similar phenomenon applies to class $b$.

The \textit{distortion region} is defined as the union of all such probability mass regions that lie outside their corresponding cells and enter those of other classes, i.e.\ $\widehat{A_1OB_1} \cup \widehat{A_2OB_2}$. The distortion region reflects the geometric mismatch between class-conditional embedding distributions and the decision regions they induce, and directly leads to misclassification and prediction bias under dual imbalance, causing increases in the false positive rate (FPR) and the false negative rate (FNR).

It is important to distinguish the two sides of this mismatch, because they call for different remedies. The extent of a class's probability mass region, measured by $\theta^\text{learned}_i$, is a property the model can shape during training. By contrast, the room the class is granted is not an independent quantity. The apex angle of an IVC is fixed by the number of classes alone. The cell a prototype actually governs, however, follows from where the prototypes settle. Under dual imbalance the crowded, hard classes end up confined to cells narrower than their IVCs, so a class whose mass would fit comfortably inside an IVC can still spill out of the narrower cell it is assigned. The distortion region is therefore driven not by the extent of the embeddings alone, but by the mismatch between the cone a class carries and the room it is given.

Regardless of whether a low $\kappa_i$ is caused by sample scarcity (frequency imbalance) or pattern heterogeneity (difficulty imbalance), the resulting geometric consequence on the unit hypersphere is the same: the LVC extends beyond the IVC, thereby forming a distortion region. The adaptive margin therefore penalizes only the extent to which the LVC exceeds the IVC, without requiring the underlying cause of this deviation to be identified. In this way, MARGIN unifies both types of imbalance under a single geometric objective, namely concentration alignment, thereby avoiding the trade-offs associated with multi-objective optimization.

Therefore, the core challenge is not merely the class imbalance itself, but the misalignment between the geometry of embedding distributions and the induced decision regions. An effective learning strategy should explicitly constrain the embedding distribution to reduce the distortion region.

To this end, we introduce a margin constraint into the Cosine Softmax Loss, enforcing that each sample of class $i$ remains within an angular distance $m_i$ from its corresponding prototype. This imposes an upper bound on the angular support of the class-conditional distribution, effectively contracting its probability mass region (i.e., reducing $\theta^\text{learned}_i$). As a result, intra-class embeddings become more compact and better aligned with their decision regions, thereby reducing the distortion region.

However, a fixed margin contradicts the heterogeneous dispersion induced by dual imbalance (e.g., $\theta^\text{learned}_a > \theta^\text{learned}_b$). Classes with highly dispersed probability mass regions require stronger constraints to sharpen the decision boundary, while more concentrated classes benefit from looser margins to maintain stable optimization. This reveals a fundamental limitation of static margin designs~\cite{Lampert_2023,10.1609/aaai.v39i19.34261}.

To address this, we propose a dynamic margin mechanism that adapts $m_i$ according to the concentration of each class during training. The objective of MARGIN can then be stated at two levels. Directly, the margin contracts the LVC so that the probability mass region (e.g., $\widehat{A_1OA_2}$) no longer overruns the cell (e.g., $\widehat{Q_1P_aQ_2}$). Ultimately, the $|\mathcal{Y}|$ Voronoi cells tile the hypersphere exactly. Confining every class's mass within the angular budget of an IVC therefore forces the prototypes apart until the ideal tiling is recovered. At that point the Learned Voronoi Cone of every class has contracted to the Ideal one: $\theta^\text{learned}_i \;\longrightarrow\; \theta^\text{ideal},
\quad \forall i \in \mathcal{Y}$.
Reaching this limit removes the prediction bias induced by imbalance, since the decision regions it produces are exactly the symmetric, unbiased partition that balanced data would have induced. Similarly, the above analysis and solution can be naturally extended to the multi-class setting.

Note that the margin is defined against the constant $\theta^\text{ideal}$ instead of the current $\theta^\text{learned}_i$. This is deliberate. The extent of a class's mass region, $\theta^\text{learned}_i$, is precisely the quantity that dual imbalance has already distorted, and adopting it as the target would merely reproduce the asymmetric partition the imbalanced data has induced instead of correcting it. Anchoring instead on $\theta^\text{ideal}$, which depends only on the number of classes, supplies the scale of the desired symmetric partition and leaves the prototype configuration free to converge to it.

\section{Methodology}
\label{sec:methodology}

\subsection{Feature Extraction and Normalization}

Given a code sample $\mathcal{X}$, we employ CodeT5 as the transformer backbone to extract a 768-dimensional semantic vector $\mathbf{e}_n \in \mathbb{R}^{768}$, where $n$ indexes the samples. We adopt CodeT5 following prior work, as it is widely used in code representation tasks; our ablation studies confirm that the observed gains stem from MARGIN and not from the backbone.

To enforce a geometrically interpretable space, we apply $\ell_2$-normalization to both feature embeddings and classifier prototypes.

\subsection{Approximation of the Concentration Parameter}

To quantify each class's concentration on the hypersphere, we estimate the vMF parameter $\kappa_i$ via Maximum Likelihood Estimation (MLE). Given the normalized embeddings of class $i$, we first compute the empirical mean vector $\boldsymbol{\mu}_i$ to get its norm and the mean resultant length.

The MLE of $\kappa_i$ involves a ratio of modified Bessel functions, which has no closed-form inverse and usually requires iterative optimization. To simplify, we use the high-dimensional approximation proposed in~\cite{vmf}.

\begin{equation}
    \label{form:banerjee-kappa}
    \kappa_i \approx \frac{||\boldsymbol{\mu}_i||\,(d-\boldsymbol{\mu}_i^2)}{1-\boldsymbol{\mu}_i^2}
\end{equation}

Here, $d$ is the embedding dimensionality and $||\boldsymbol{\mu}_i||$ is the mean resultant length of class $i$. The approximation is accurate in high dimensions; with CodeT5 producing 768-dimensional embeddings, it is both empirically justified and widely used in prior work.

\subsection{Aligning the Voronoi Cone}

To describe the angular uncertainty of each class on the
hypersphere, we estimate the apex angle of the Learned Voronoi Cone of each class from its underlying vMF distribution.

$\theta^{\text{learned}}_i$ denotes the \emph{full} apex angle, that is, the angle between two opposite generatrices of the cone. The colatitude of the $\alpha$-confidence cap is therefore the half-apex angle $\theta^{\text{learned}}_i/2$. Under a vMF distribution for class $i$, the confidence level $\alpha$ satisfies

\begin{equation}
\alpha=
\frac{
\int_{0}^{\theta^\text{learned}_i/2}
(\sin\theta)^{d-2} e^{\kappa_i \cos\theta}\, d\theta
}{
\int_{0}^{\pi}
(\sin\theta)^{d-2} e^{\kappa_i \cos\theta}\, d\theta
}
\end{equation}

In practice, we set $\alpha = 95\%$, and the corresponding apex angle of the Learned Voronoi Cone is given by

\begin{equation}
\theta^{\text{learned}}_i
=
2\arccos\left(
F_{i}^{-1}(1-\alpha)
\right)
\end{equation}

where $F_{i}^{-1}$ denotes the inverse marginal cumulative
distribution function of the cosine $\tilde{\boldsymbol{\mu}}_i^\top \mathbf{\tilde{e}}$ under the vMF distribution of class $i$. The argument is $1-\alpha$ because a larger angle corresponds to a smaller cosine. As no closed-form solution exists, we adopt an efficient
approximation. Given the concentration parameter $\kappa_i$ of class
$i$, we use it directly as the effective concentration.

Based on the effective concentration, the apex angle of the Learned Voronoi Cone
is approximated as

\begin{equation}
\label{form:learned-apex-approx}
\theta_i^{\text{learned}}
=
2\sqrt{
\frac{
\chi^2_{\alpha}(d-1)
}{
\kappa_i
}
}
\end{equation}

\noindent where $\chi^2_{\alpha}(d-1)$ denotes the $\alpha$-quantile of the
chi-square distribution with $d-1$ degrees of freedom, and the right-hand side is the closed form computed by the algorithm. The approximation is valid for $\kappa_i\gg d-1$ and its error is bounded in Appendix~\ref{secA1}. This follows
from the asymptotic property of vMF distributions that, under high
concentration, local angular deviations can be approximated by a
Gaussian distribution on the tangent space~\cite{pmlr-v32-gopal14,mettes2019hyperspherical}.
The full derivation, together with a Taylor-expansion error analysis, is given in
Appendix~\ref{secA1}. That analysis separates the error of the tangent-space limit
from the error of the small-angle inversion and shows that
Eq.~(\ref{form:learned-apex-approx}) is a \emph{conservative upper bound} on the exact
apex angle. Because the margin penalizes the excess of the
Learned cone over the Ideal cone, this conservative bias strengthens the intended
regularization and never weakens it. Appendix~\ref{secA1} also shows analytically
that the apex angle is insensitive to the confidence level $\alpha$ at the embedding
dimension used here, which supports treating $\alpha$ as a fixed statistical
constant.

To constrain each class's probability mass region within its cell, MARGIN uses an adaptive angular margin that penalizes the extent to which the Learned Voronoi Cone exceeds the Ideal Voronoi Cone. Additionally, since the Voronoi cone is symmetric, the apex angle must be halved accordingly. The margin of the inward-shrinking angle is defined with a non-negativity constraint:

\begin{equation}
\label{form:learned-apex-angle}
m_i
=
\text{max}
\left(
\frac{
\theta_i^{\text{learned}}
-
\theta^{\text{ideal}}
}{2},0
\right)
\end{equation}

In practice, however, the regularization force gradually weakens in the later stages of training, as $\theta^\text{learned}_i$ approaches $\theta^\text{ideal}$. This reduces the optimization pressure on the embedding geometry and may cause training to stagnate. To prevent this late-stage weakening of regularization, MARGIN continues to contract the cone toward the tightest Learned Voronoi apex angle currently attained by any class, $\theta^\text{learned}_\text{min} = \min_i \theta^\text{learned}_i$. This additional target maintains sufficient optimization pressure throughout the later training stage, allowing the embedding geometry to continue being effectively regularized and preventing premature convergence. Therefore:

\begin{equation}
m_i
=
\text{max}
\left(
\frac{
\theta_i^{\text{learned}}
-
\theta^{\text{ideal}}
}{2},
\frac{
\theta^\text{learned}_i -\theta^\text{learned}_\text{min}
}{2}
\right)
\end{equation}

The excess of a Learned Voronoi Cone over its Ideal Voronoi Cone marks where
two classes may overlap, so penalizing this excess improves
hyperspherical separability and reduces class interference.

\subsection{Concentration-Aware Logit Scaling}

To adapt classifier confidence to class compactness, we adjust the
logit scaling according to the convergence status of each class.

The concentration parameter $\kappa_i$ reflects the compactness of class
representations on the hypersphere. A larger $\kappa_i$ indicates that
class $i$ is well optimized with low intra-class variance; assigning large
logits in this case may lead to over-confident predictions and may dominate
gradient updates. Therefore, we assign smaller scaling factors to such
classes to suppress their influence. In contrast, a smaller $\kappa_i$
indicates a more dispersed distribution and insufficient convergence,
for which we increase the scaling factors to sharpen logits and amplify
gradient signals.

To achieve this behavior while avoiding the instability of explicit
inverse mappings, we adopt a \emph{soft ranking-based strategy} in the
log-concentration space, assigning larger weights to lower-concentration classes. Specifically, we compute:

\begin{equation}
r_i = |\mathcal{Y}| \cdot \mathrm{softmax}\!\left(-\frac{\log \kappa_i}{|\mathcal{Y}|}\right)
\end{equation}

where $|\mathcal{Y}|$ is the number of classes, ensuring that $r_i > 0$ with a unit mean.

Finally, the class-wise scaling factors are:
\begin{equation}
s_i = s_0 \cdot {r}_i
\end{equation}
where $s_0$ is a base scaling factor.

This formulation enables geometry-aware and difficulty-adaptive gradient
reallocation, enhancing under-converged classes while preventing
well-optimized ones from dominating training.

\subsection{Objective Function}

Combining the adaptive margin and the concentration-aware scaling factor, the objective for a sample of class $i$ is:

\begin{equation}
\ell_i
=
- \log
\frac{\exp\!\big(s_i\cos(\theta_{i} + m_i)\big)}
{\exp(s_i\cos(\theta_{i} + m_i)) + \sum_{j \neq i} \exp\!\big(s_j \cos \theta_j\big)}
\end{equation}

Note that in very rare, extreme cases the angle $\theta_i + m_i$ may exceed $\pi$ under low concentration, where the cosine is no longer monotonic. We therefore add monotonicity protection in numerical processing and replace the target-class cosine with a numerically robust, monotonicity-preserving form. Writing $x=\cos\theta_i$, the corrected target similarity is
\[
\resizebox{0.98\columnwidth}{!}{$\displaystyle
\phi(x)=
\begin{cases}
x\cos m_i-\sqrt{\max(1-x^2,0)}\,\sin m_i, & x>-\cos m_i,\\[2pt]
-1, & x\le -\cos m_i,
\end{cases}$}
\]
and $\phi(\cos\theta_i)$ is used in place of $\cos(\theta_i+m_i)$ for the target class only. The two branches are continuous with matching first derivatives at $x=-\cos m_i$, and the $\max$ term prevents numerical error when $1-x^2<0$ under floating-point rounding. Appendix~\ref{secA2} proves the $C^1$ property, establishes that the corrected logit is monotone non-increasing in the angular distance, and details the floating-point safeguards. The correction is inactive whenever $\theta_i+m_i\le\pi$, so it leaves the standard formulation unchanged over the entire practically relevant range.

During training progress, we calculate the adaptive parameters $s_i$ and $m_i$ per epoch to calibrate the manifold structure.

\subsection{Inference at Test Time}

At the evaluation stage, instead of directly using classifier weights as prototypes, we construct data-driven prototypes from training embeddings to better capture the empirical geometry of the feature space. Specifically, for each class $i \in \mathcal{Y}$, we compute the geometric median of its normalized training embeddings and project it onto the unit hypersphere:

\begin{equation}
\mathbf{p}_i
=
\arg\min_{\mathbf{p}\in\mathbb{R}^{d}}
\sum_{j:\,y_j=i}
\left\|
\mathbf{p}-\tilde{\mathbf{e}}_j
\right\|_2,
\quad
\tilde{\mathbf{p}}_i=\frac{\mathbf{p}_i}{\|\mathbf{p}_i\|_2}
\end{equation}

Compared to weight-based prototypes, geometric medians are more robust to outliers and better reflect class-wise embedding distributions~\cite{minsker2024geometric}.

Inference is performed via nearest-prototype matching in angular space:
\begin{equation}
\hat y
=
\arg\max_{y\in\mathcal Y}
\tilde{\mathbf e}_{\text{test}}^\top
\tilde{\mathbf p}_y
\end{equation}

This induces a Voronoi partition on $\mathbb S^{d-1}$ in which each sample is assigned to the angular region of its closest prototype, i.e., to the cell of that prototype. Driving the Learned Voronoi Cones toward the Ideal ones during training, as described in Section~\ref{sec:motivation-analysis}, is what makes this partition match the class structure at inference time.

Furthermore, statistical analysis and training dynamics indicate that the classifier weight prototypes and inference-time geometric median prototypes converge toward one another as training approaches the Neural Collapse regime. Consequently, the Voronoi cells induced by these two types of prototypes become highly consistent and statistically equivalent in their classification behavior. A detailed analysis of their convergence and equivalence is provided in Appendix~\ref{secA3} from a mathematical perspective. This result provides theoretical justification for replacing classifier weight prototypes with geometric median prototypes during inference.

\subsection{Training Procedure}

For implementation clarity, we provide a reference algorithm in Algorithm~\ref{alg:margin}. The model is initialized with $m_i=0$ and $s_i=s_0$, reducing to plain cosine softmax in the first epoch before geometric calibration takes effect. Class statistics are accumulated from detached embeddings produced during the standard forward pass, requiring neither an additional data pass nor gradient computation. At the end of each epoch, calibration is performed under \texttt{no\_grad}, and the updated parameters are applied in the subsequent epoch.

\begin{algorithm}[t]
\caption{MARGIN: training and inference}
\label{alg:margin}
\begin{algorithmic}[1]
\Require encoder $f_\theta$; class weights $\{\mathbf w_i\}$; base scale $s_0$; confidence $\alpha$; dimension $d$; number of classes $C$
\State $m_i\gets 0$,\ \ $s_i\gets s_0$ \quad for all $i$ \Comment{warm-up epoch is plain cosine softmax}
\State $q\gets\chi^2_{\alpha}(d-1)$;\ \ $\theta^{\text{ideal}}\gets\arccos\!\big(-1/(C-1)\big)$
\For{each epoch}
  \State $\mathcal E_i\gets\emptyset$ for all $i$ \Comment{per-class feature buffers}
  \For{each mini-batch $(\mathcal X,\mathbf y)$}
    \State $\mathbf e\gets f_\theta(\mathcal X)$;\ \ $\tilde{\mathbf e}\gets\mathbf e/\|\mathbf e\|_2$
    \State $\cos\theta_j\gets\tilde{\mathbf e}^\top\tilde{\mathbf w}_j$ for all $j$, clamped to $[-1,\,1]$
    \State $z_y\gets s_y\,\phi(\cos\theta_y;\,m_y)$;\ \ $z_j\gets s_j\cos\theta_j$ for $j\neq y$
    \State $\mathcal L\gets\mathrm{CrossEntropy}(\mathbf z,\mathbf y)$;\ \ update $f_\theta,\{\mathbf w_i\}$
    \State $\mathcal E_y\gets\mathcal E_y\cup\{\mathrm{detach}(\mathbf e)\}$ for each sample label $y$
  \EndFor
  \Statex \hspace{\algorithmicindent}\textit{Geometric calibration (no gradient)}
  \For{each class $i$}
    \State $\mathbf X_i\gets$ row-wise $\ell_2$ normalize of $\mathcal E_i$
    \State $\tilde{\boldsymbol\mu}_i\gets\mathrm{normalize}\big(\mathrm{mean}(\mathbf X_i)\big)$
    \State $\tilde{\mathbf p}_i\gets\mathrm{Weiszfeld}(\mathbf X_i)$ \Comment{re-normalized each iteration}
    \State $\bar R_i\gets\mathrm{mean}\big(\mathbf X_i\tilde{\boldsymbol\mu}_i\big)$, clamped to $(0,1)$
    \State $\kappa_i\gets\bar R_i(d-\bar R_i^2)/(1-\bar R_i^2)$, floored at $1$
    \State $\theta_i^{\text{learned}}\gets 2\sqrt{q/\kappa_i}$
  \EndFor
  \State $\theta^{\text{learned}}_{\min}\gets\min_i\theta_i^{\text{learned}}$
  \State $\mathbf r\gets C\cdot\mathrm{softmax}\big(-\log\boldsymbol\kappa/C\big)$
  \For{each class $i$}
    \State $m_i\gets\max\Big\{\tfrac{1}{2}\big(\theta_i^{\text{learned}}-\theta^{\text{ideal}}\big),\ \tfrac{1}{2}\big(\theta_i^{\text{learned}}-\theta^{\text{learned}}_{\min}\big),\ 0\Big\}$
    \State $s_i\gets s_0\,r_i$
  \EndFor
  \State validate with $\hat y=\arg\max_{y}\ \tilde{\mathbf e}^\top\tilde{\mathbf p}_y$;\ \ retain the best checkpoint
\EndFor
\State \Return best checkpoint and its prototypes $\{\tilde{\mathbf p}_i\}$ for test-time matching
\end{algorithmic}
\end{algorithm}

\section{Experiments}
\label{sec:experiments}

All experiments were conducted on a single NVIDIA RTX 5880 Ada GPU (48 GB) with CUDA 12.4. All methods were evaluated under identical experimental settings unless otherwise specified. For both the baseline and ablation experiments, each experiment was repeated five times with different random seeds to validate the effectiveness of the proposed method and reduce the effects of random fluctuations. We report the mean and standard deviation across the five runs. To evaluate the effectiveness of MARGIN thoroughly, we formulate the following research questions:

\begin{itemize}
    \item \textbf{RQ1:} How effective is the proposed method compared with baseline approaches?
    \item \textbf{RQ2:} How do the proposed adaptive mechanisms contribute to performance?
    \item \textbf{RQ3:} How sensitive is MARGIN to different hyperparameter settings?
    \item \textbf{RQ4:} What is the computational cost of MARGIN?
\end{itemize}

\subsection{Datasets}

\begin{figure*}\centering \includegraphics[width=0.8\textwidth]{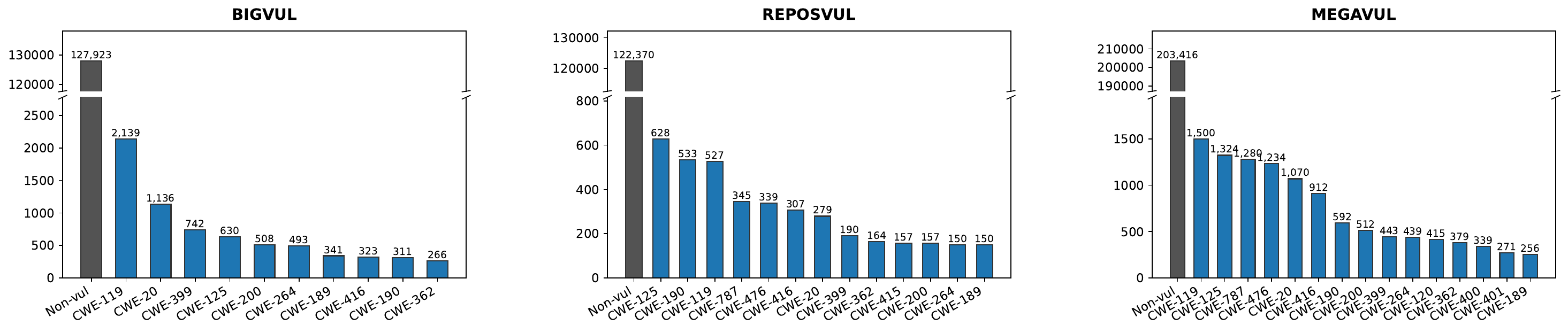} \caption{\textbf{Class distribution across labels.} In terms of frequency, non-vulnerable samples dominate by a wide margin, while the remaining CWE vulnerability samples show a long-tailed distribution. Overall, the label distribution is highly imbalanced.}
\label{fig:dataset-statistics}
\end{figure*}

We evaluate MARGIN on three widely used public vulnerability detection datasets: \textbf{BigVul}~\cite{bigvul}, \textbf{MegaVul}~\cite{megavul}, and \textbf{ReposVul}~\cite{reposvul}. These datasets, collected from open-source repositories, CVE and National Vulnerability Database (NVD) reports, and vulnerability-fixing commits, vary in scale and CWE distribution. They are highly imbalanced, with non-vulnerable samples dominating, making them realistic benchmarks for imbalanced vulnerability detection.

BigVul and MegaVul are large-scale C/C++ datasets, while ReposVul emphasizes higher data quality through automated vulnerability disentanglement and dependency-aware filtering. These datasets vary in scale, data sources, and CWE distributions. BigVul and MegaVul primarily cover C/C++ vulnerabilities, whereas ReposVul emphasizes data quality through automated vulnerability disentanglement and dependency-aware filtering. All three show marked class imbalance, with non-vulnerable samples dominating.

All datasets are processed at the function level, with each sample representing a single function. We use the functions identified by BigVul and MegaVul and the preprocessed function-level version released by ReposVul. Each function is assigned its corresponding CWE label based on CVE/NVD records, while non-vulnerable functions are labeled \texttt{Non-Vul}. This unified representation ensures compatibility with function-level baselines.

Following prior work, we perform standard preprocessing and data splitting. To construct a high-quality and statistically meaningful dataset, we select more than ten CWE categories with highly imbalanced class frequencies and manifest intra-class heterogeneity to reflect the realistic characteristics of vulnerability data for training and evaluation. The resulting datasets exhibit coefficients of variation ranging from 3.13 to 3.78, indicating considerable class imbalance. We split each dataset into training, validation, and test sets at a ratio of 8:1:1.

\begin{table}[t]
\centering
\caption{Statistics of vulnerability detection datasets. IR denotes the imbalance ratio between the Non-Vul and all CWE-* classes, and CV denotes the coefficient of variation of the per-class sample counts.}
\label{tab:dataset-statistics}
\setlength{\tabcolsep}{3pt}
\begin{tabular}{@{}lrrrrr@{}}
\toprule
Dataset & Samples & IR & CWEs & CV & Source \\
\midrule
BigVul   & 134,812  & 18.57 & 10 & 3.13 & 348 Projects. \\
ReposVul & 126,296  & 31.17 & 13 & 3.61 & 1,491 Projects. \\
MegaVul  & 214,382 & 18.55 & 15 & 3.78 & 992 Projects. \\
\botrule
\end{tabular}
\end{table}

\subsection{Baselines}

To evaluate the effectiveness of MARGIN thoroughly for CWE-based single-label multi-class vulnerability classification, we compare the proposed method against existing classical and state-of-the-art representative baselines from three categories of similar work in vulnerability detection, including graph-based methods, program analysis--based approaches, and recent representation learning models. Specifically, the compared vulnerability detection baselines fall into three groups. The first comprises \textbf{ReVeal} (2021, TSE)~\cite{reveal} and \textbf{LIVABLE} (2024, TSE)~\cite{livable}, which use graph neural networks for structural vulnerability modeling. The second comprises \textbf{$\mu$VulDeePecker} (2019, DSC)~\cite{muvuldeepecker} and \textbf{SySeVR} (2022, DSC)~\cite{sysevr}, which exploit vulnerability-aware program analysis and code slicing. The third comprises \textbf{HCL-VC} (2024, EMNLP)~\cite{hclvc}, \textbf{MoEVD} (2025, FSE)~\cite{moevd}, and \textbf{CLeVeR} (2025, ACL)~\cite{clever}, which represent recent advances in CWE-aware representation learning and vulnerability classification. In addition to vulnerability detection methods, we also include representative long-tailed and class-imbalanced learning approaches, including \textbf{Focal Loss}~\cite{focalloss}, \textbf{Class-Balanced Loss (CBL)}~\cite{Cui_2019}, and \textbf{LDAM Loss}~\cite{ldam}, to assess whether the performance gains of MARGIN arise from its geometry-aware design and not from conventional imbalance-aware loss functions. For fair comparison, we follow the hyperparameter settings recommended in the original papers. Models are reproduced based on publicly released code or reported configurations. For the imbalance-learning baselines, all relevant hyperparameters follow the best configurations recommended in their original papers; for framework-based baselines, we use the same backbone network as MARGIN (CodeT5) to ensure a fair comparison.

\subsection{Metrics}
\label{sec:metrics}

We evaluate MARGIN from two complementary perspectives: (i) \textbf{Binary Vulnerability Detection} and (ii) \textbf{Fine-grained CWE Classification}. Given the severe class imbalance, we adopt imbalance-robust metrics.

For binary detection, we collapse all CWE categories into a single CWE positive class (\textit{Vulnerable}) and treat \texttt{Non-Vul} as the negative class. We report MCC, F1-score, Recall, and Precision. The predicted category is $\hat y=\arg\max_{y\in\mathcal Y}\tilde{\mathbf e}_{\text{test}}^\top\tilde{\mathbf p}_y$, and a sample is declared \emph{vulnerable} if and only if $\hat y$ is any \texttt{CWE-*} category, and \emph{non-vulnerable} if and only if $\hat y=\texttt{Non-Vul}$.

For multi-CWE classification, we follow a one-vs-all (OvA) scheme over all classes to compute per-class MCC, F1-score, Recall, and Precision. To specifically assess vulnerability-type recognition, we exclude the \texttt{Non-Vul} class and compute macro-averaged metrics over all CWE classes, referred to as \textbf{CWE-Macro} metrics.

This design evaluates the method from two complementary angles, namely its ability to separate vulnerable from non-vulnerable code and its fine-grained discrimination among vulnerability types.

\section{Discussion}

\subsection{Overall Performance Comparison (RQ1)}

We compare MARGIN with representative vulnerability detection methods
on BigVul, MegaVul, and ReposVul. The results in Table~\ref{tab:baseline-comparison}
show that MARGIN consistently outperforms existing approaches in both
binary vulnerability detection and CWE classification.

On BigVul, MARGIN achieves the best overall performance, obtaining the
highest MCC and F1-scores for binary detection. It also significantly improves
Macro-MCC and Macro-F1 for CWE classification compared with previous methods.
This indicates that the proposed adaptive geometric constraints not only improve
the separation between vulnerable and non-vulnerable samples, but also enhance
discrimination among different CWE categories.

On MegaVul, which contains a larger and more challenging vulnerability
distribution, MARGIN maintains consistent improvements over all
baselines. Compared with recent structure-aware methods such as LIVABLE, HCL-VC,
and CLeVeR, it achieves better detection performance while providing stronger
CWE classification capability. These results show that the proposed method
is robust to large-scale data and severe class imbalance.

On ReposVul, which reflects practical repository-level scenarios,
MARGIN also achieves the best overall performance. Compared with existing
methods, it provides a better balance between recall and precision in vulnerability
detection and obtains higher macro-level classification performance, suggesting
better generalization to real-world vulnerability patterns.

Existing slice-based approaches, such as SySeVR and $\mu$VulDeePecker, tend to
achieve high recall but suffer from low precision, resulting in excessive false
positives. Structure-aware methods reduce this problem by incorporating code
structure information, but still struggle with rare CWE categories. In contrast,
MARGIN learns more discriminative vulnerability representations through
adaptive geometric optimization, improving both detection reliability and
fine-grained CWE recognition.

Beyond vulnerability-specific detectors, a comparison against imbalance-aware
loss functions clarifies where the gains of MARGIN originate. Although
Focal Loss, CBL, and LDAM are designed explicitly to counteract skewed label
distributions, they remain behind MARGIN on every dataset and every
metric. On BigVul, the strongest of the three is LDAM, reaching an MCC of 0.8281
and an F1 of 83.28\%, still 0.0702 and 7.02 points below MARGIN, while
its Macro-F1 of 56.56\% trails by 15.80 points. The gap persists on the more
challenging datasets. On ReposVul, LDAM obtains the best F1 of the three,
35.95\% against 45.91\% for MARGIN, a margin of 9.96 points. On
MegaVul, the best of the three is CBL with an F1 of 47.16\% and a Macro-F1 of
26.19\%, against 54.36\% and 35.73\% for MARGIN. In particular,
MARGIN also attains these results with a markedly smaller variance,
with an F1 standard deviation of 0.70--0.89 against 1.47--2.47 for this family.

Two observations suggest that such losses reduce the symptom and not the
cause of class imbalance. First, they largely move the operating point along the
precision--recall trade-off instead of improving the underlying representation.
On ReposVul, Focal Loss and CBL achieve the highest precision of all compared
methods (53.32\% and 52.55\%), but only at a recall of 24.73\% and 21.06\%,
whereas MARGIN reaches a comparable precision (47.96\%) with nearly
twice the recall (44.03\%). The opposite imbalance appears on BigVul, where LDAM
retains a recall close to that of MARGIN (91.19\% vs. 92.93\%) yet
loses 11.18 points of precision (76.63\% vs. 87.81\%). Second, the deficit
widens on CWE-level metrics, which weight every category equally: on BigVul and
MegaVul the Macro-MCC margins over the strongest baseline of this family (0.1502 and 0.0984) exceed the corresponding binary MCC margins (0.0702 and 0.0840).
This indicates that reweighting the objective does not by itself produce the
well-separated, fine-grained class regions required for CWE recognition, whereas
the geometry-aware constraints of MARGIN do.

Statistical analysis further confirms the effectiveness of MARGIN, all reported improvements in F1 and Macro-F1 being statistically significant ($p<0.05$ across all datasets and baselines);
the improvements over Focal Loss, CBL, and LDAM in particular are significant at
$p<0.01$ on all three datasets.
Overall, the results show that MARGIN provides more robust
vulnerability analysis by improving feature discrimination, particularly under
highly imbalanced CWE distributions.

\begin{table*}[t]
\centering
\caption{Baseline comparison. Results are averaged over five runs. $p$-values are computed on F1 and Macro-F1 scores against MARGIN.}
\label{tab:baseline-comparison}
\resizebox{\ifdim\width>\linewidth\linewidth\else\width\fi}{!}{
\begin{tabular}{llcccccccc}
\toprule
\multirow{2}{*}{Dataset}
& \multirow{2}{*}{Method}
& \multicolumn{4}{c}{Vulnerability Detection (Vul/Non-Vul Binary)}
& \multicolumn{4}{c}{Vulnerability Classification (CWE-Macro)} \\
\cmidrule(lr){3-6} \cmidrule(lr){7-10}
&
& MCC & F1(\%) ($p$-value) & R(\%) & P(\%)
& Macro-MCC & Macro-F1(\%) ($p$-value) & Macro-R(\%) & Macro-P(\%) \\
\midrule

\multirow{11}{*}{\textbf{BigVul}}
 & ReVeal                 & 0.7888$\pm$0.0196 & 79.74$\pm$2.83 ($6.9{\times}10^{-4}$) & 87.37 & 73.33 & 0.2506$\pm$0.0268 & 21.76$\pm$1.89 ($3.7{\times}10^{-9}$) & 30.88 & 34.31 \\

 & LIVABLE                & 0.8369$\pm$0.0258 & 84.48$\pm$1.47 ($2.1{\times}10^{-4}$) & 88.10 & 81.15 & 0.6191$\pm$0.0139 & 60.86$\pm$2.25 ($8.6{\times}10^{-5}$) & 60.87 & 65.95 \\

\cmidrule(lr){2-10}
 & $\mu$VulDeePecker      & 0.6930$\pm$0.0274 & 68.58$\pm$1.66 ($3.4{\times}10^{-7}$) & 93.61 & 54.11 & 0.4153$\pm$0.0172 & 39.54$\pm$2.57 ($1.2{\times}10^{-6}$) & 42.91 & 46.61 \\

 & SySeVR                 & 0.6554$\pm$0.0237 & 63.94$\pm$1.84 ($3.4{\times}10^{-7}$) & \textbf{95.36} & 48.10 & 0.4403$\pm$0.0221 & 42.61$\pm$1.58 ($8.6{\times}10^{-9}$) & 52.39 & 42.28 \\

\cmidrule(lr){2-10}
 & HCL-VC                 & 0.8464$\pm$0.0183 & 85.42$\pm$1.02 ($4.2{\times}10^{-5}$) & 85.05 & 85.80 & 0.6637$\pm$0.0247 & 66.17$\pm$1.54 ($1.6{\times}10^{-4}$) & 68.62 & 65.17 \\

 & MoEVD                  & 0.8184$\pm$0.0219 & 82.74$\pm$2.78 ($2.6{\times}10^{-3}$) & 85.92 & 79.78 & 0.3499$\pm$0.0291 & 33.93$\pm$1.37 ($2.0{\times}10^{-10}$) & 38.40 & 35.24 \\

 & CLeVeR                 & 0.8430$\pm$0.0288 & 85.06$\pm$1.59 ($6.4{\times}10^{-4}$) & 88.82 & 81.60 & 0.6409$\pm$0.0279 & 63.76$\pm$1.96 ($1.4{\times}10^{-4}$) & 66.11 & 63.48 \\

\cmidrule(lr){2-10}
 & Focal Loss             & 0.8164$\pm$0.0185 & 82.56$\pm$1.70 ($1.3{\times}10^{-4}$) & 83.73 & 81.42 & 0.5446$\pm$0.0190 & 53.86$\pm$1.65 ($3.1{\times}10^{-7}$) & 57.43 & 53.64 \\

 & CBL    & 0.7167$\pm$0.0185 & 72.25$\pm$1.70 ($1.2{\times}10^{-6}$) & 61.73 & 87.08 & 0.4401$\pm$0.0190 & 42.09$\pm$1.65 ($1.4{\times}10^{-8}$) & 40.13 & 53.68 \\

 & LDAM                   & 0.8281$\pm$0.0123 & 83.28$\pm$1.47 ($7.2{\times}10^{-5}$) & 91.19 & 76.63 & 0.5730$\pm$0.0112 & 56.56$\pm$1.39 ($1.3{\times}10^{-7}$) & 63.40 & 53.47 \\

\cmidrule(lr){2-10}
 & \textbf{MARGIN} & \textbf{0.8983}$\pm$0.0083 & \textbf{90.30}$\pm$0.79 (--) & 92.93 & \textbf{87.81} & \textbf{0.7232}$\pm$0.0109 & \textbf{72.36}$\pm$0.95 (--) & \textbf{74.55} & \textbf{70.55} \\
\midrule

\multirow{11}{*}{\textbf{ReposVul}}
 & ReVeal                 & 0.3869$\pm$0.0174 & 39.88$\pm$2.66 ($5.2{\times}10^{-3}$) & 52.42 & 32.19 & 0.3213$\pm$0.0286 & 31.29$\pm$1.83 ($5.1{\times}10^{-3}$) & 39.15 & 27.76 \\

 & LIVABLE                & 0.3715$\pm$0.0226 & 37.90$\pm$1.92 ($2.1{\times}10^{-4}$) & 54.20 & 29.14 & 0.3120$\pm$0.0149 & 29.80$\pm$2.68 ($8.0{\times}10^{-3}$) & 41.48 & 25.15 \\

\cmidrule(lr){2-10}
 & $\mu$VulDeePecker      & 0.4050$\pm$0.0287 & 42.01$\pm$1.53 ($2.2{\times}10^{-3}$) & 51.15 & 35.64 & 0.3146$\pm$0.0252 & 30.84$\pm$1.68 ($2.0{\times}10^{-3}$) & 36.99 & 28.02 \\

 & SySeVR                 & 0.2203$\pm$0.0249 & 16.77$\pm$1.81 ($8.5{\times}10^{-8}$) & \textbf{82.19} & 9.34 & 0.1931$\pm$0.0158 & 13.72$\pm$2.34 ($9.3{\times}10^{-6}$) & \textbf{53.71} & 8.62 \\

\cmidrule(lr){2-10}
 & HCL-VC                 & 0.3471$\pm$0.0263 & 33.85$\pm$1.76 ($1.1{\times}10^{-5}$) & 60.81 & 23.45 & 0.3074$\pm$0.0172 & 28.56$\pm$2.31 ($1.8{\times}10^{-3}$) & 44.65 & 24.12 \\

 & MoEVD                  & 0.4105$\pm$0.0195 & 42.66$\pm$2.48 ($4.0{\times}10^{-2}$) & 39.95 & 45.77 & 0.2980$\pm$0.0277 & 29.59$\pm$1.97 ($1.6{\times}10^{-3}$) & 28.03 & 32.62 \\

 & CLeVeR                 & 0.3895$\pm$0.0254 & 40.24$\pm$1.68 ($5.2{\times}10^{-4}$) & 51.91 & 32.85 & 0.3231$\pm$0.0184 & 31.47$\pm$1.03 ($2.0{\times}10^{-4}$) & 40.16 & 27.15 \\

\cmidrule(lr){2-10}
 & Focal Loss             & 0.3340$\pm$0.0187 & 33.79$\pm$2.14 ($5.2{\times}10^{-5}$) & 24.73 & \textbf{53.32} & 0.3148$\pm$0.0173 & 30.23$\pm$2.05 ($3.3{\times}10^{-3}$) & 28.96 & \textbf{38.29} \\

 & CBL    & 0.3000$\pm$0.0231 & 30.07$\pm$1.96 ($6.1{\times}10^{-6}$) & 21.06 & 52.55 & 0.3119$\pm$0.0208 & 29.59$\pm$1.72 ($7.1{\times}10^{-4}$) & 30.42 & 37.50 \\

 & LDAM                   & 0.3441$\pm$0.0176 & 35.95$\pm$2.32 ($2.5{\times}10^{-4}$) & 29.23 & 46.68 & 0.2913$\pm$0.0191 & 28.36$\pm$2.17 ($1.1{\times}10^{-3}$) & 28.22 & 32.61 \\

\cmidrule(lr){2-10}
 & \textbf{MARGIN} & \textbf{0.4414}$\pm$0.0084 & \textbf{45.91}$\pm$0.89 (--) & 44.03 & 47.96 & \textbf{0.3540}$\pm$0.0088 & \textbf{35.36}$\pm$0.73 (--) & 35.22 & 36.26 \\
\midrule

\multirow{11}{*}{\textbf{MegaVul}}
 & ReVeal                 & 0.3280$\pm$0.0162 & 31.88$\pm$2.28 ($7.1{\times}10^{-6}$) & 72.84 & 20.40 & 0.1742$\pm$0.0236 & 15.17$\pm$1.44 ($1.8{\times}10^{-7}$) & 32.21 & 10.97 \\

 & LIVABLE                & 0.4749$\pm$0.0295 & 50.18$\pm$1.42 ($1.2{\times}10^{-3}$) & 50.32 & 50.05 & 0.3280$\pm$0.0173 & 32.68$\pm$1.01 ($8.1{\times}10^{-4}$) & 32.00 & \textbf{34.74} \\

\cmidrule(lr){2-10}
 & $\mu$VulDeePecker      & 0.2502$\pm$0.0211 & 22.94$\pm$2.92 ($7.8{\times}10^{-6}$) & 80.58 & 13.37 & 0.0656$\pm$0.0274 & 5.35$\pm$1.69 ($1.2{\times}10^{-7}$) & 16.55 & 4.95 \\

 & SySeVR                 & 0.2732$\pm$0.0243 & 24.30$\pm$1.76 ($1.9{\times}10^{-7}$) & \textbf{84.78} & 14.19 & 0.1224$\pm$0.0194 & 9.53$\pm$2.53 ($7.1{\times}10^{-6}$) & 30.73 & 6.37 \\

\cmidrule(lr){2-10}
 & HCL-VC                 & 0.4306$\pm$0.0256 & 45.64$\pm$1.85 ($1.6{\times}10^{-4}$) & 56.97 & 38.06 & 0.2574$\pm$0.0196 & 25.20$\pm$2.41 ($3.3{\times}10^{-4}$) & 30.71 & 23.15 \\

 & MoEVD                  & 0.4653$\pm$0.0179 & 49.31$\pm$2.67 ($1.2{\times}10^{-2}$) & 50.32 & 48.34 & 0.2898$\pm$0.0258 & 28.85$\pm$1.86 ($5.2{\times}10^{-4}$) & 29.44 & 29.76 \\

 & CLeVeR                 & 0.4825$\pm$0.0267 & 50.82$\pm$1.58 ($4.7{\times}10^{-3}$) & 49.59 & 52.11 & 0.3043$\pm$0.0212 & 30.21$\pm$2.69 ($8.5{\times}10^{-3}$) & 29.57 & 32.67 \\

\cmidrule(lr){2-10}
 & Focal Loss             & 0.4409$\pm$0.0162 & 46.27$\pm$2.24 ($7.2{\times}10^{-4}$) & 52.55 & 41.33 & 0.2487$\pm$0.0174 & 23.56$\pm$2.38 ($1.6{\times}10^{-4}$) & 31.04 & 22.62 \\

 & CBL    & 0.4435$\pm$0.0208 & 47.16$\pm$1.55 ($1.2{\times}10^{-4}$) & 47.80 & 46.53 & 0.2656$\pm$0.0197 & 26.19$\pm$2.41 ($5.2{\times}10^{-4}$) & 28.36 & 26.49 \\

 & LDAM                   & 0.3934$\pm$0.0183 & 42.26$\pm$2.47 ($2.1{\times}10^{-4}$) & 35.46 & \textbf{52.28} & 0.2548$\pm$0.0168 & 24.27$\pm$2.35 ($2.0{\times}10^{-4}$) & 27.97 & 26.93 \\

\cmidrule(lr){2-10}
 & \textbf{MARGIN} & \textbf{0.5275}$\pm$0.0079 & \textbf{54.36}$\pm$0.70 (--) & 60.87 & 49.54 & \textbf{0.3640}$\pm$0.0084 & \textbf{35.73}$\pm$0.70 (--) & \textbf{41.53} & 31.99 \\
\botrule
\end{tabular}
}
\end{table*}

\subsection{Contribution of Adaptive Mechanisms (RQ2)}

\begin{table*}[t]
\centering
\caption{Ablation study of key components in MARGIN.}
\label{tab:ablation}
\resizebox{\ifdim\width>\linewidth\linewidth\else\width\fi}{!}{
\begin{tabular}{lcccccccc}
\toprule
\multirow{2}{*}{Variant}
& \multicolumn{4}{c}{Vulnerability Detection (Vul/Non-Vul Binary)}
& \multicolumn{4}{c}{Vulnerability Classification (CWE-Macro)} \\
\cmidrule(lr){2-5} \cmidrule(lr){6-9}
& MCC & F1(\%) & R(\%) & P(\%)
& Macro-MCC & Macro-F1(\%) & Macro-R(\%) & Macro-P(\%) \\
\midrule

Cosine Softmax Only
& 0.8594$\pm$0.0154 & 86.56$\pm$1.21 & 83.16 & \textbf{90.24} & 0.6403$\pm$0.0202 & 63.78$\pm$1.80 & 60.67 & 68.57 \\

\midrule
w/o Adaptive Margin ($m_i$)
& 0.8447$\pm$0.0280 & 85.25$\pm$2.90 & 87.66 & 82.97 & 0.6892$\pm$0.0190 & 68.79$\pm$2.00 & 71.01 & 67.76 \\

w/o Adaptive Logit Scaling ($s_i$)
& 0.8663$\pm$0.0250 & 87.27$\pm$2.60 & 85.05 & 89.60 & 0.6966$\pm$0.0180 & 69.38$\pm$1.90 & 66.67 & 73.83 \\

\midrule
w/ Classifier Weight Prototypes
& 0.8712$\pm$0.0400 & 87.71$\pm$2.60 & 90.70 & 84.91 & 0.7078$\pm$0.0360 & 70.75$\pm$2.70 & 68.33 & \textbf{73.88} \\

\midrule
\textbf{MARGIN (Full)}
& \textbf{0.8983}$\pm$0.0083 & \textbf{90.30}$\pm$0.79 & \textbf{92.93} & 87.81 & \textbf{0.7232}$\pm$0.0109 & \textbf{72.36}$\pm$0.95 & \textbf{74.55} & 70.55 \\

\botrule
\end{tabular}
}
\end{table*}

To investigate the contribution of each adaptive component in MARGIN,
we perform ablation studies on the BigVul dataset by removing the adaptive margin and the adaptive
logit scaling separately. The results are shown in Table~\ref{tab:ablation}.

Removing the adaptive margin (\textit{w/o Adaptive Margin}) leads to performance
degradation in both vulnerability detection and CWE classification. The reduced
precision indicates that the decision boundary becomes less calibrated, resulting
in more false positives. This shows that the adaptive margin effectively
optimizes inter-class distances and improves feature discrimination. The decline
in Macro-MCC and Macro-F1 further verifies its importance in maintaining balanced
performance across different CWE categories.

Removing the adaptive logit scaling (\textit{w/o Adaptive Logit Scaling}) has a smaller impact on binary detection but causes more noticeable
degradation in CWE classification. This suggests that logit scaling improves
class-wise calibration, especially for minority CWE categories in imbalanced
datasets, by enhancing the discrimination among vulnerability types.

We further compare geometric median prototypes with classifier weight prototypes during inference. Although classifier weight prototypes achieve competitive performance, they show larger variations across runs. In contrast, geometric median prototypes yield higher performance and smaller standard deviations, showing their stronger robustness to outliers and greater stability in prototype estimation.

The full MARGIN model achieves the best performance on both tasks. Its
improvement over the Cosine Softmax baseline indicates that the gains mainly come
from the proposed adaptive geometric constraints and not from the backbone alone.
The improvements in macro-level metrics further show its effectiveness in
handling long-tailed CWE distributions through better intra-class compactness and
inter-class separation.

Overall, the ablation results confirm that the two adaptive mechanisms are
complementary: adaptive margin refines feature geometry, while adaptive logit
scaling improves prediction calibration. Together, they enable MARGIN to
achieve robust vulnerability detection and fine-grained CWE classification under
class imbalance.

\subsection{Hyperparameter Sensitivity Analysis (RQ3)}

\begin{table}[t]
\centering
\caption{Sensitivity analysis of the base scaling $s_0$ on BigVul. This sweep uses a single seed across all 12 settings so that the trend is not confounded by seed variation.}
\label{tab:hyperparameter-sensitivity}
\begin{tabular}{@{}c|cc@{}}
\toprule
$s_0$ & Binary F1 (\%) & CWE-Macro F1 (\%) \\
\midrule
5  & 88.91 & 67.27 \\
10 & 85.98 & 63.78 \\
15 & 89.25 & 71.07 \\
20 & \textbf{89.66} & \textbf{72.91} \\
25 & 89.25 & 71.07 \\
30 & 87.45 & 67.82 \\
35 & 88.35 & 66.21 \\
40 & 87.14 & 68.18 \\
45 & 86.86 & 70.10 \\
50 & 85.98 & 68.40 \\
55 & 85.25 & 68.79 \\
60 & 85.00 & 69.01 \\
\botrule
\end{tabular}
\end{table}

Although MARGIN incorporates adaptive margins and class-specific adjustments,
it introduces only one explicit hyperparameter, the base scaling
factor $s_0$, which controls the strength of geometric modulation in the
hyperspherical feature space. The confidence level $\alpha$ is not a tunable
hyperparameter. It is a fixed statistical constant that defines the probability
mass covered by the class cap, and Appendix~\ref{secA1} shows analytically that at
$d=768$ a change of one percentage point in $\alpha$ moves the apex angle by only
about $0.24\%$, while sweeping $\alpha$ over $[0.80,0.99]$ changes it by $3.8\%$ in
total. The same analysis shows that the elasticity grows sharply as $\alpha$
approaches one, which is why the conventional value $\alpha=0.95$ is retained.

We investigate the sensitivity to $s_0$ on BigVul, and the results are presented
in Table~\ref{tab:hyperparameter-sensitivity}. The performance follows a
non-monotonic trend, where moderate scaling values consistently achieve better
results than extremely small or large values. The best performance is obtained
when $s_0=20$, achieving the highest Binary F1 and CWE-Macro F1, indicating that
an appropriate scaling factor can effectively balance feature compactness and
class separation.

When $s_0$ is relatively small, the geometric modulation is insufficient to
provide strong class discrimination, resulting in weaker performance, especially
for CWE classification. As $s_0$ increases into the moderate range 15--25, the
model benefits from enhanced inter-class separation while maintaining stable
decision boundaries. However, excessively large scaling factors introduce overly
aggressive margin adjustments, which may disturb the learned feature geometry
and lead to performance fluctuations.

Overall, the sensitivity analysis shows that MARGIN is relatively
robust to the choice of $s_0$, with stable performance within a reasonable range.
The results also validate that the proposed adaptive geometric mechanism does not
rely on precise hyperparameter tuning, while an appropriate scaling factor can
further improve vulnerability detection and CWE classification.

\subsection{Computation Costs of MARGIN (RQ4)}

Due to differences in hardware, pretraining strategies, implementations, and data preprocessing, direct efficiency comparisons with prior work may be unfair. We therefore report only the efficiency of the proposed method.

We adopt the approximation of $\kappa$ given in Eq.~(\ref{form:banerjee-kappa}), which has constant time complexity without additional computational overhead. Also, computing class prototypes by the Weiszfeld algorithm with fixed dimension and iterations has linear time complexity $O(n)$, which matches the per-iteration complexity of training. Therefore, the prototype computation does not increase the overall time complexity or delay convergence, and is considered acceptable for practical use.

These statistics require no additional full-dataset forward pass. As shown in Algorithm~\ref{alg:margin}, normalized embeddings are accumulated per class during the standard training loop and used at the end of each epoch to estimate $\kappa_i$ and the geometric-median prototypes $\tilde{\mathbf p}_i$. The inference prototypes are computed offline after training. Thus, the reported training latency includes the per-epoch calibration overhead, with no additional inference-time pass.

MARGIN's computational cost in training and inference is largely determined by the backbone encoder. The adaptive margin and scaling factor mechanisms introduce additional constraints, which empirically improve gradient signals and accelerate convergence. Using CodeT5 as the backbone, the average training latency is \textbf{167--200 ms} per sample, and the inference latency is \textbf{111--125 ms} per batch, measured with a batch size of 16 under our experimental settings. On BigVul, the model typically converges within \textbf{10--20 epochs}, after which gains are marginal. Peak GPU memory usage during training is around 12 GB, well within industrial capacity.

Given these measurements, the computational cost of MARGIN is considered acceptable for industrial deployment under typical resource and latency constraints.

\section{Case Study}
\label{sec:case-study}

We use the BigVul dataset as a case study. Motivated by the preceding analysis, we examine training dynamics and key metrics to show why the model remains effective under class imbalance.

\begin{figure*}
	\centering
	\includegraphics[width=\textwidth]{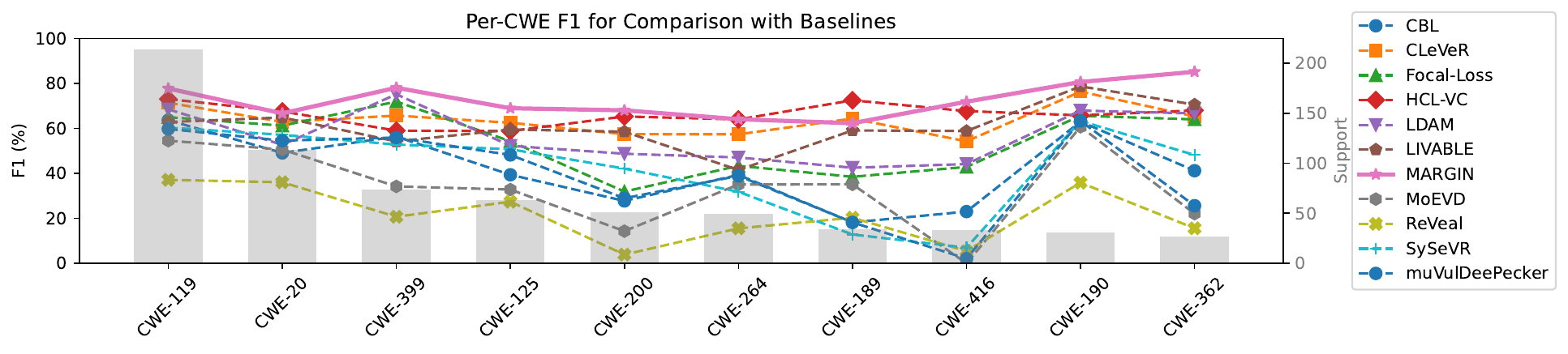}
	\caption{\textbf{Per-CWE F1 Performance Comparison with Baselines.} Per-class F1-scores and support for MARGIN and baseline methods across different CWE categories. The bar chart indicates the number of samples per class (support), while the line plots represent F1-scores. MARGIN is highlighted with a solid green line to emphasize its performance.}
	\label{fig:per-cwe-metrics}
\end{figure*}

\begin{figure*}
	\centering
	\includegraphics[width=\textwidth]{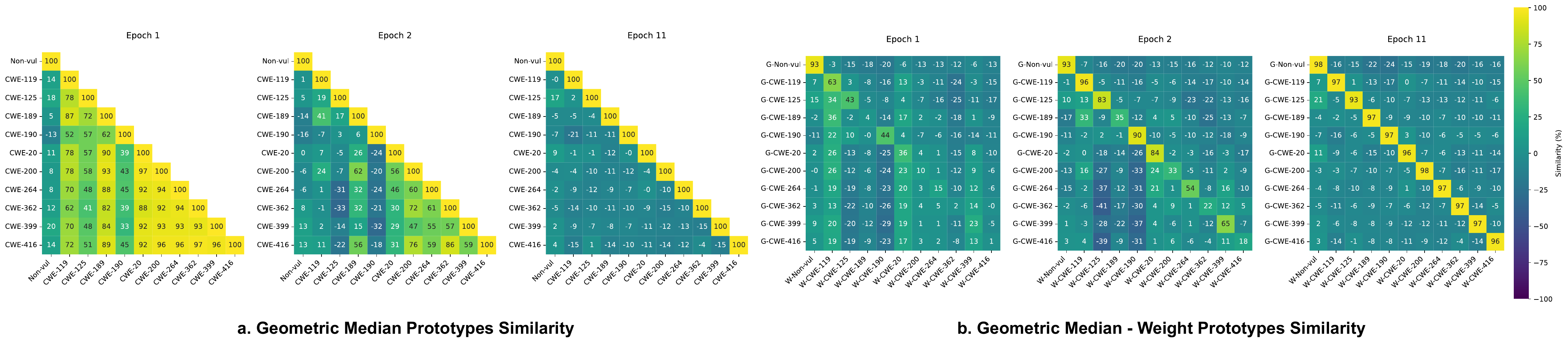}
	\caption{\textbf{Prototype Convergence During Training.} Guided by the adaptive margin mechanism, the geometric median prototypes are not only mutually separated but also aligned with the weight prototypes, suggesting that the embedding distributions of different classes gradually form a clear, separable, and stable structure.}
	\label{fig:prototype-convergence}
\end{figure*}

\begin{figure*}
	\centering
	\begin{minipage}[b]{0.48\textwidth}
		\centering
		\includegraphics[width=\textwidth]{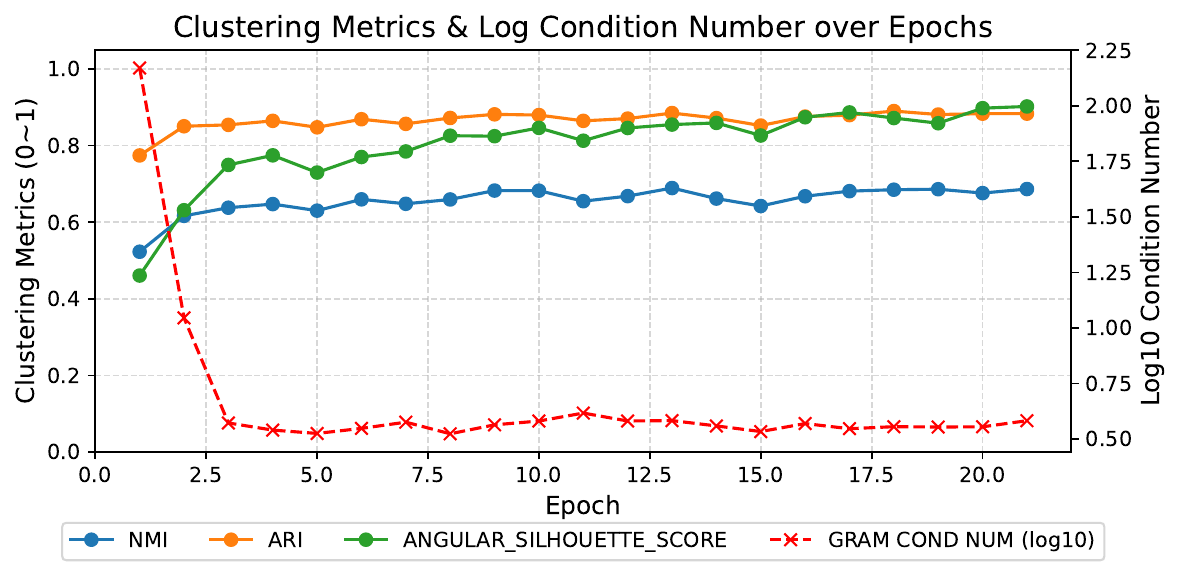}
		\par\vspace{2pt}
		{\small (a) Clustering and ETF dynamics}
	\end{minipage}\hfill
	\begin{minipage}[b]{0.48\textwidth}
		\centering
		\includegraphics[width=\textwidth]{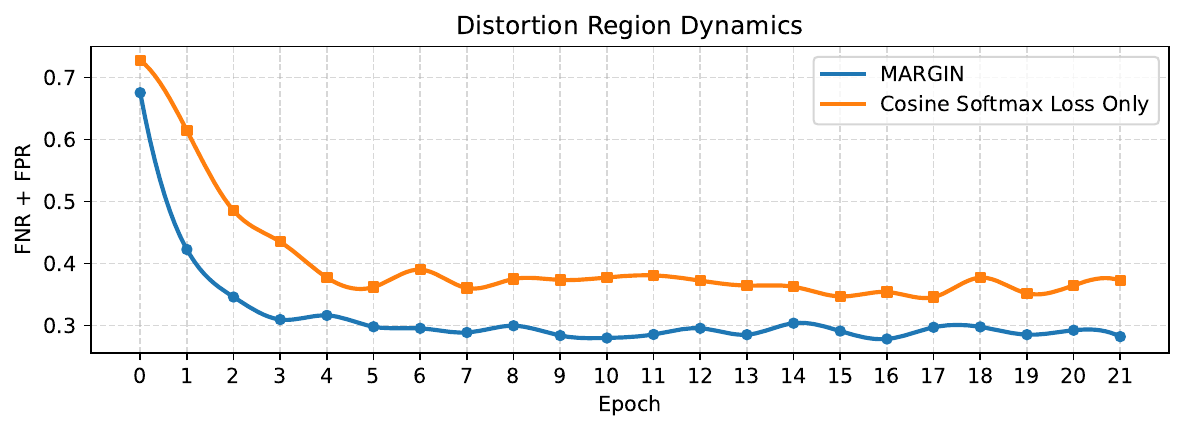}
		\par\vspace{2pt}
		{\small (b) Distortion region dynamics}
	\end{minipage}
	\caption{\textbf{Training dynamics of the embedding geometry.} (a) Evolution of normalized mutual information (NMI), adjusted Rand index (ARI), angular silhouette, and the Gram condition number of the geometric median prototypes during training. (b) Comparison of the global macro FNR and FPR across training epochs, with and without the adaptive mechanism.}
	\label{fig:training-dynamics}
\end{figure*}

\subsection{Per-CWE Performance}

To evaluate the classification performance of the proposed method across different classes, we present the per-CWE F1 performance, examining how classes with varying frequencies are classified. This allows the robustness of the proposed method under extreme class imbalance to be verified.

Figure~\ref{fig:per-cwe-metrics} illustrates the per-class F1-scores in relation to the support (number of samples) for various CWE types, providing a detailed view of how each method copes with class imbalance. A clear pattern emerges from the visualization: while baseline methods show considerable volatility that correlates with sample size, the proposed MARGIN method (depicted as a solid green line) shows remarkable stability and robustness across all classes. Unlike the baselines, MARGIN consistently achieves high F1-scores, generally above 60\%. This shows minimal sensitivity to fluctuations in support, as indicated by the grey bars.

The advantages of MARGIN are particularly pronounced in long-tailed settings, where data scarcity presents a significant challenge. For example, in CWE-416, which shows extreme imbalance with very few samples, most competing methods such as ReVeal, MoEVD, and $\mu$VulDeePecker suffer dramatic performance degradation, with F1-scores approaching zero. In contrast, MARGIN maintains a robust F1 of approximately 60\%, indicating its ability to learn discriminative features even in severely underrepresented classes. Moreover, while methods like CLeVeR and LIVABLE achieve competitive peaks in high-support classes (e.g., CWE-190), they display notable instability in medium-support categories such as CWE-200 and CWE-189. MARGIN, however, preserves a steady performance across the spectrum, showing that it does not merely overfit to head classes but generalizes effectively to both frequent and rare vulnerability types.

\subsection{Concentration Dynamics and Stability Analysis}
\label{sec:kappa-dynamics}

\begin{figure*}[t]
\centering
\begin{minipage}[b]{0.48\textwidth}
\centering
\includegraphics[width=\textwidth]{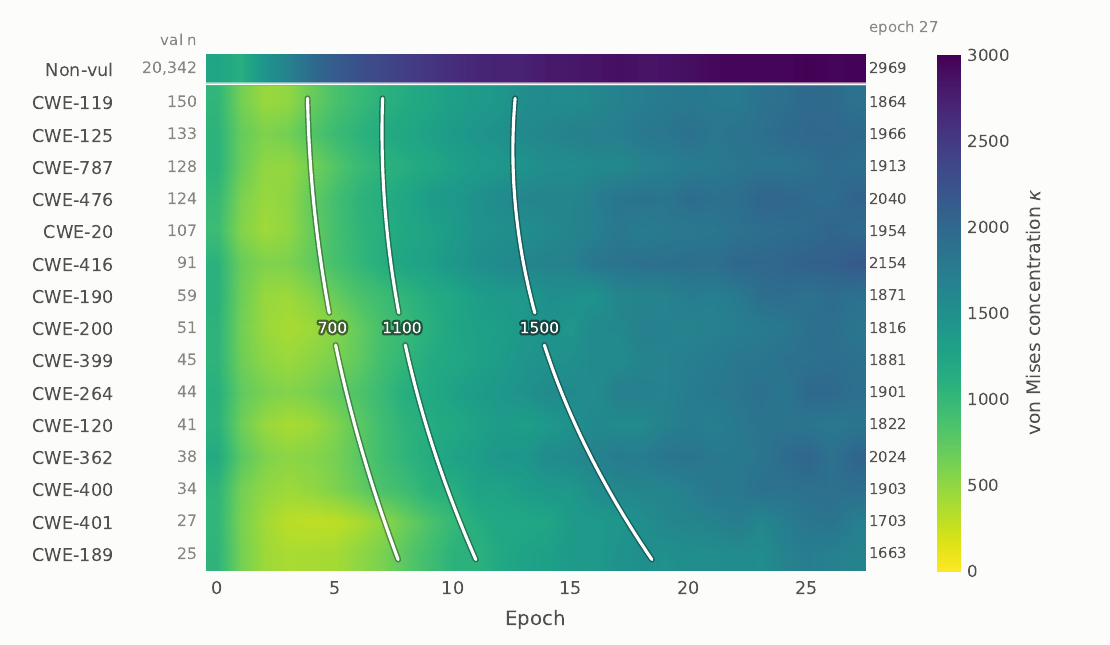}
\par\vspace{2pt}
{\small (a) Class-wise $\kappa$ across epochs}
\end{minipage}\hfill
\begin{minipage}[b]{0.48\textwidth}
\centering
\includegraphics[width=\textwidth]{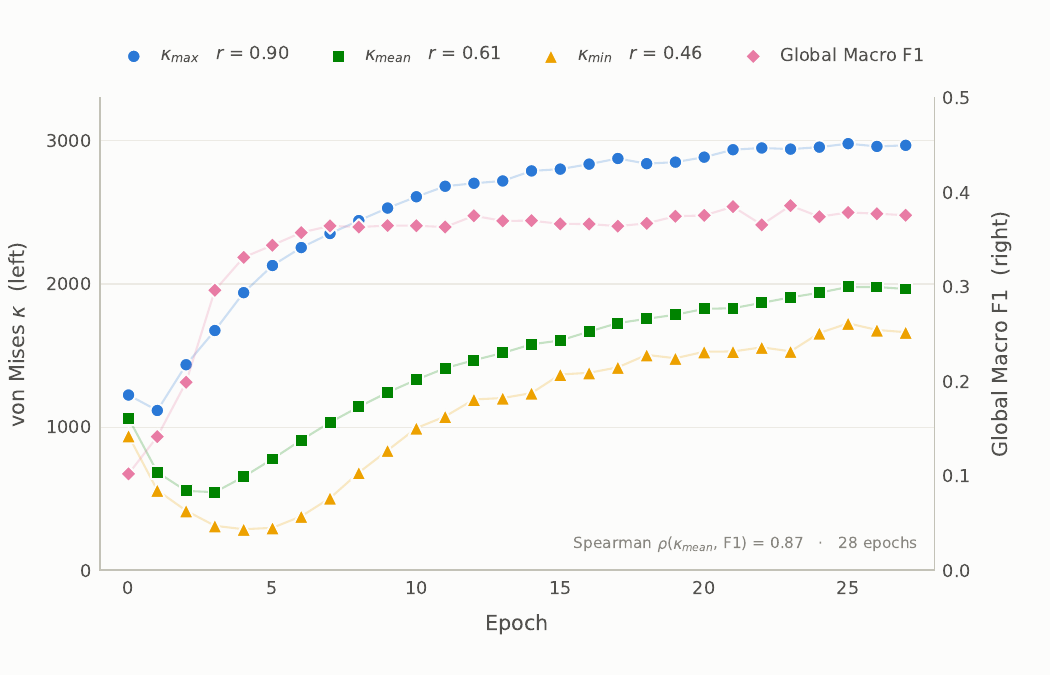}
\par\vspace{2pt}
{\small (b) Class-wise Macro-F1 versus $\kappa$}
\end{minipage}
\caption{\textbf{Concentration dynamics on MegaVul.} (a) Heatmap of the per-class concentration parameter $\kappa_i$ over training epochs. (b) Scatter plot of per-class Macro-F1 against $\kappa_i$ across epochs, with the global Macro-F1 shown on the secondary axis.}
\label{fig:kappa-dynamics}
\end{figure*}

The adaptive mechanisms act on the estimated concentration $\kappa_i$, so it is natural to ask whether these estimates are stable enough to drive the geometry, in particular for categories with limited support. We examine this on MegaVul, which has the largest number of retained categories and therefore the widest spread of class sizes. Figure~\ref{fig:kappa-dynamics} reports two complementary views.

To better understand the training dynamics of MARGIN, we examine the evolution of class-wise concentration parameters and their relationship with model performance in Figure~\ref{fig:kappa-dynamics}(a). The $\kappa$ heatmap reveals distinct valley-like patterns during the early warm-up stage (Epochs 1--10), with lower-frequency classes showing slower adaptation. After Epoch 10, concentration levels increase progressively, while the contours become smoother across classes, reflecting faster convergence of majority classes and slower adaptation of minority classes. By approximately Epoch 20, the concentration gradients become more consistent, suggesting that class representations are gradually reaching a more balanced geometric configuration.

This evolution is further supported by the trajectories of the maximum, mean, and minimum $\kappa$ values over 28 epochs, alongside the global Macro-F1 scores in Figure~\ref{fig:kappa-dynamics}(b). Although $\kappa_{\min}$ and $\kappa_{\mathrm{mean}}$ undergo initial adjustments, all three statistics subsequently increase smoothly and stabilize, without pronounced epoch-to-epoch fluctuations, even for the minimum concentration. These observations suggest that MARGIN progressively adapts to class-specific representation characteristics while maintaining stable concentration estimation across classes, including minority classes.

Two design choices further limit the influence of estimation noise. First, the prototypes are geometric medians, whose breakdown point of $0.5$ makes them resistant to outliers and to sparse sampling~\cite{minsker2024geometric}. Second, the class-wise scaling factor $s_i$ reallocates gradient magnitude according to convergence status, which damps the influence of classes whose concentration estimate is still noisy. Appendix~\ref{secA1} adds a first-order bound, showing that a relative perturbation $\delta\kappa/\kappa$ induces only half as large a relative perturbation in the apex angle, and an absolute margin perturbation of at most $\tfrac14\theta_i^{\text{learned}}\,\delta\kappa/\kappa$.

\subsection{Structure Convergence}

Figure~\ref{fig:training-dynamics}(a) tracks the first 20 epochs, with clustering metrics (NMI, ARI, angular silhouette) on the left axis and the condition number of the prototype Gram matrix on the right. The clustering metrics rise rapidly and the condition number drops sharply in the early epochs, with all of them plateauing around epoch 7, indicating fast convergence to a discriminative embedding space. Figure~\ref{fig:prototype-convergence} visualizes the dynamics of the embedding distribution during optimization, as measured by the geometric median prototype. It indicates that the trained embedding space converges toward the discriminative structure required for category classification. It also confirms the equivalence and consistency of the two prototype families, as discussed in Appendix~\ref{secA3}. In particular, the condition number decreases sharply to approximately 1, suggesting a transition from an ill-conditioned, highly correlated prototype geometry to a stable equicorrelated regime with greatly reduced redundancy among class representations. This reflects a progressive decorrelation of prototype directions and improved numerical stability of the embedding space. These results show that the learned prototypes rapidly self-organize into a stable and well-distributed configuration on the hypersphere. Rather than collapsing, the embedding space evolves toward a uniformly coupled geometric structure, where classes maintain moderate and approximately homogeneous angular relationships, supporting effective separation under the adaptive margin mechanism.

\subsection{Distortion Elimination}

As discussed in Section~\ref{sec:motivation-analysis}, the distortion region refers to the phenomenon where the probability mass of one class intrudes into the Voronoi cells of other classes, directly leading to increases in global macro FNR and FPR. Our adaptive margin mechanism explicitly aims to reduce these distortion regions and thereby improve classification accuracy. Figure~\ref{fig:training-dynamics}(b) illustrates the dynamics of the global macro FNR and FPR across training epochs. For comparison, we also show results obtained without the adaptive mechanism, using only the Cosine Softmax Loss, highlighting the effectiveness of the adaptive margin in reducing distortion regions.

Due to computational constraints, precisely quantifying the distortion region on the unit hypersphere is infeasible. Although the overlap between class-conditional distributions can in principle be measured with standard divergence measures~\cite{lin1991divergence,endres2003new}, reliably estimating them in high-dimensional hyperspherical space is impractical. As discussed earlier, the distortion region is closely related to FPR and FNR, allowing its size to be estimated indirectly from these metrics for each class. We compare MARGIN with a variant that removes the adaptive mechanism, relying solely on the Cosine Softmax Loss (Figure~\ref{fig:training-dynamics}(b)). Even at later training stages, the Cosine Softmax Loss shows relatively high FPR and FNR, whereas MARGIN significantly reduces both. This suggests that the adaptive mechanism reduces distortion across classes and improves alignment within each region, thereby enhancing robustness under class imbalance.

\section{Threats to Validity}

\subsection{External Validity.} This study is limited to C/C++ and function-level vulnerability detection for two main reasons. First, most existing baselines are primarily implemented and evaluated on C/C++, with inconsistent support for other languages, making C/C++ the only setting that enables reliable and fair comparison. Second, large-scale standardized cross-language benchmarks for vulnerability detection remain limited, leaving cross-language generalizability, particularly for language-agnostic vulnerability patterns, an open question for future work. All three datasets and baselines are processed at the function level without interprocedural call context, calling preconditions, or external dependency information. We therefore restrict our claims to function-level CWE classification on repository-derived datasets. Extending the input to incorporate cross-language data and interprocedural or repository-level context is beyond the scope of the present study and remains an important direction for future work.

\subsection{Construct Validity.} The experiments use CodeT5 as the encoder, with the primary objective of investigating the effectiveness of the proposed geometric regularization for imbalanced vulnerability detection. MARGIN operates on $\ell_2$-normalized embeddings, and the experiments focus on validating the proposed geometric mechanisms within this setting. A systematic evaluation across different encoder architectures is a separate research direction beyond the scope of this study and is left for future work.

\section{Conclusion and Future Work}

We study vulnerability detection under realistic, highly imbalanced CWE distributions from an embedding geometry perspective. By regularizing the hyperspherical structure of representations, the proposed method enhances class separability and decision boundary stability, achieving consistent gains over strong baselines while providing a geometrically interpretable, imbalance-aware framework. The experiments also provide empirical evidence for this approach, supporting the rationale and effectiveness of the inductive bias underlying the vMF-based modeling.

Future work will explore richer code representations that incorporate structural and semantic information like causality to improve discriminability for rare and diverse CWE categories. We will also investigate cross-language and cross-project generalization, as well as dynamic imbalance-aware learning that jointly considers class frequency, learning difficulty, and evolving representation geometry. Finally, we plan to extend the proposed geometric perspective to more challenging settings, such as hierarchical and open-set vulnerability detection.

\appendices
\onecolumn
\numberwithin{equation}{section}

\section{Derivation of the Learned Voronoi Cone Apex Angle}
\label{secA1}

This appendix derives the approximation used for the apex angle $\theta_i^{\mathrm{learned}}$ in Eq.~(\ref{form:learned-apex-approx}), analyzes its error, and studies its sensitivity to the confidence level $\alpha$. Throughout, the analysis is carried out on the unit hypersphere $\mathbb{S}^{d-1}$ embedded in $\mathbb{R}^d$, and all angles are in radians.

\subsection{The von~Mises--Fisher model}

For each class $i$, the normalized embeddings $\mathbf{x}\in\mathbb{S}^{d-1}$ are modeled by a vMF distribution with mean direction $\tilde{\boldsymbol{\mu}}_i$ and concentration $\kappa_i\ge0$~\cite{vmf}, with normalizing constant
\begin{equation}
C_d(\kappa)=\frac{\kappa^{d/2-1}}{(2\pi)^{d/2}\,I_{d/2-1}(\kappa)},
\label{eq:vmf-norm}
\end{equation}
where $I_{\nu}$ is the modified Bessel function of the first kind of order $\nu$.

\subsection{Exact definition of the apex angle}

Rotate coordinates so that $\tilde{\boldsymbol{\mu}}_i$ is the north pole and write $\mathbf{x}=\tilde{\boldsymbol{\mu}}_i\cos\theta+\boldsymbol{\xi}\sin\theta$ with $\boldsymbol{\xi}\in\mathbb{S}^{d-2}$, so that $\theta\in[0,\pi]$ is the colatitude. The uniform surface measure factorizes as
\begin{equation}
d\sigma_{d-1}(\mathbf{x})
=\omega_{d-2}\,(\sin\theta)^{d-2}\,d\theta\,d\sigma_{d-2}(\boldsymbol{\xi}),
\qquad
\omega_{d-2}=\frac{2\pi^{(d-1)/2}}{\Gamma\!\left(\tfrac{d-1}{2}\right)} .
\end{equation}
Because the density depends on $\mathbf{x}$ only through $\cos\theta$, integrating out $\boldsymbol{\xi}$ gives the marginal density of $\Theta_i=\arccos(\tilde{\boldsymbol{\mu}}_i^\top\mathbf{x})$,
\begin{equation}
f_{\Theta_i}(\theta)
=\frac{(\sin\theta)^{d-2}\,e^{\kappa_i\cos\theta}}
       {\int_{0}^{\pi}(\sin\theta)^{d-2}\,e^{\kappa_i\cos\theta}\,d\theta},
\qquad 0\le\theta\le\pi .
\label{eq:angle-density}
\end{equation}

The Learned Voronoi Cone has its apex at the origin and its base at the spherical cap centered at $\tilde{\boldsymbol{\mu}}_i$ carrying a fraction $\alpha$ of the class mass. Let $\theta_i^{\mathrm{learned}}$ denote the \emph{full} apex angle, that is, the angle between two opposite generatrices. The colatitude of the cap is therefore the half-apex angle $\theta_i^{\mathrm{learned}}/2$, and
\begin{equation}
\alpha
=P\!\left(\Theta_i\le\frac{\theta_i^{\mathrm{learned}}}{2}\right)
=\frac{\int_{0}^{\theta_i^{\mathrm{learned}}/2}(\sin\theta)^{d-2}\,e^{\kappa_i\cos\theta}\,d\theta}
       {\int_{0}^{\pi}(\sin\theta)^{d-2}\,e^{\kappa_i\cos\theta}\,d\theta} .
\label{eq:apex-exact}
\end{equation}
In terms of the cosine $t=\cos\theta$, whose marginal density is proportional to $e^{\kappa_i t}(1-t^2)^{(d-3)/2}$ on $[-1,1]$, this reads $\theta_i^{\mathrm{learned}}=2\arccos(F_i^{-1}(1-\alpha))$, where $F_i$ is the cumulative distribution function of $t$.

\subsection{Exact representation via modified Bessel functions}

The normalization integral in \eqref{eq:angle-density} is known in closed form,
\begin{equation}
\int_{0}^{\pi} e^{\kappa\cos\theta}\,(\sin\theta)^{d-2}\,d\theta
=\frac{\sqrt{\pi}\,\Gamma\!\left(\tfrac{d-1}{2}\right)}{(\kappa/2)^{d/2-1}}\;I_{d/2-1}(\kappa),
\label{eq:bessel-int}
\end{equation}
where \eqref{eq:bessel-int} is the standard integral representation of the modified Bessel function~\cite{abramowitz1964handbook}. Hence the denominator of \eqref{eq:angle-density} equals $1/[C_d(\kappa)\,\omega_{d-2}]$, recovering \eqref{eq:vmf-norm}. The quantile involves the incomplete counterpart of \eqref{eq:bessel-int} and admits no elementary closed form, which motivates the approximation derived next.

\subsection{High-concentration limit}

Set $V=2\kappa_i(1-\cos\Theta_i)$, so that $\theta=\arccos(1-\tfrac{v}{2\kappa_i})$ and $\mathrm{d}\theta/\mathrm{d}v=1/(2\kappa_i\sin\theta)$. The density of $V$ is
\[
f_V(v)=\frac{(\sin\theta)^{d-3}\,e^{\kappa_i\cos\theta}}{2\kappa_i\,Z_{d,\kappa_i}},
\qquad
Z_{d,\kappa_i}=\int_{0}^{\pi}(\sin\theta)^{d-2}\,e^{\kappa_i\cos\theta}\,d\theta .
\]
Substituting $\cos\theta=1-\tfrac{v}{2\kappa_i}$ and $\sin\theta=\sqrt{\tfrac{v}{\kappa_i}-\tfrac{v^{2}}{4\kappa_i^{2}}}$ and letting $\kappa_i\to\infty$ at fixed $v$ gives
\[
f_V(v)\;\longrightarrow\;
\frac{v^{(d-3)/2}\,e^{-v/2}}{2^{(d-1)/2}\,\Gamma\!\left(\tfrac{d-1}{2}\right)},
\]
the density of $\chi^2_{d-1}$. Hence
\begin{equation}
2\kappa_i\,(1-\cos\Theta_i)\;\xrightarrow{\;d\;}\;\chi^2_{d-1},
\qquad
\sqrt{\kappa_i}\,\mathbf{u}_i\;\xrightarrow{\;d\;}\;\mathcal{N}\!\left(\mathbf{0},\,I_{d-1}\right),
\label{eq:chisq-limit}
\end{equation}
where $\mathbf{u}_i=\boldsymbol{\xi}\sin\theta$ is the tangent component. This is the tangent-space normal approximation of the vMF distribution~\cite{mardia2000directional,pmlr-v32-gopal14,mettes2019hyperspherical}.

\subsection{Exact quantile identity and the closed form}

Throughout this subsection every relation is an identity or carries an explicit remainder. The single approximation is stated at the end, together with its validity condition.

Introduce the chord distance
\begin{equation}
\eta=2\sin\frac{\theta}{2}\in[0,2],
\qquad
\eta^2=2(1-\cos\theta)=\left\lVert\tilde{\boldsymbol{\mu}}_i-\mathbf{x}\right\rVert^2,
\label{eq:eta}
\end{equation}
so that $V=\kappa_i\eta^2$ exactly, and \eqref{eq:chisq-limit} reads $\kappa_i\eta^2\xrightarrow{d}\chi^2_{d-1}$. The colatitude is recovered from $\eta$ by the exact inverse
\begin{equation}
\theta=2\arcsin\frac{\eta}{2}.
\label{eq:theta-inv}
\end{equation}

Let $V_\alpha$ denote the \emph{exact} $\alpha$-quantile of $V=2\kappa_i(1-\cos\Theta_i)$. Since $V$ is a strictly increasing function of $\Theta_i$ on $[0,\pi]$, quantiles are carried over exactly, so
\begin{equation}
V_\alpha=2\kappa_i\Big(1-\cos\frac{\theta_i^{\mathrm{learned}}}{2}\Big),
\qquad
\eta_\alpha=\sqrt{\frac{V_\alpha}{\kappa_i}},
\label{eq:quantile-exact}
\end{equation}
where $\eta_\alpha$ is the exact $\alpha$-quantile of $\eta$. Inverting \eqref{eq:quantile-exact} gives the exact representation
\begin{equation}
\theta_i^{\mathrm{learned}}
=2\arccos\!\Big(1-\frac{V_\alpha}{2\kappa_i}\Big)
=4\arcsin\!\Big(\frac12\sqrt{\frac{V_\alpha}{\kappa_i}}\Big).
\label{eq:exact-apex}
\end{equation}
No approximation has been made so far. The only unknown in \eqref{eq:exact-apex} is $V_\alpha$.

The inverse function is analytic at the origin, with the exact expansion
\begin{equation}
4\arcsin\frac{\eta}{2}
=2\eta+\frac{\eta^3}{12}+\frac{3\eta^5}{320}+O\!\left(\eta^7\right),
\label{eq:arcsin-series}
\end{equation}
all of whose coefficients are positive, so that
\begin{equation}
\theta_i^{\mathrm{learned}}=2\eta_\alpha+\frac{\eta_\alpha^3}{12}+O\!\left(\eta_\alpha^5\right).
\label{eq:apex-expansion}
\end{equation}

Two replacements turn \eqref{eq:apex-expansion} into a closed form, and only these two are approximations:
\begin{enumerate}
\item[(i)] the exact quantile $V_\alpha$ is replaced by the chi-square quantile $q:=\chi^2_\alpha(d-1)$, which is the limit of $V_\alpha$ by \eqref{eq:chisq-limit};
\item[(ii)] the terms $\eta_\alpha^3/12$ and higher in \eqref{eq:apex-expansion} are dropped.
\end{enumerate}
Their composition defines the closed form used by the method,
\begin{equation}
\widehat\theta_i:=2\sqrt{\frac{q}{\kappa_i}}
=2\sqrt{\frac{\chi^2_{\alpha}(d-1)}{\kappa_i}}.
\label{eq:approx-learned}
\end{equation}
Consequently, and only at this point,
\[
\theta_i^{\mathrm{learned}}\;\approx\;\widehat\theta_i,
\qquad\text{valid when }\ \kappa_i\gg d-1 ,
\]
which is Eq.~(\ref{form:learned-apex-approx}). The two errors are of opposite sign and are bounded separately in the next subsection, where their composition is shown to make $\widehat\theta_i$ an upper bound.

\subsection{Error analysis}

\paragraph{Validity regime and the two error sources.}
The limit \eqref{eq:chisq-limit} is exact as $\kappa_i\to\infty$ with $d$ fixed. A standard criterion uses the mean resultant length $\bar R_i=I_{d/2}(\kappa_i)/I_{d/2-1}(\kappa_i)$, which satisfies $1-\bar R_i=\frac{d-1}{2\kappa_i}+O(\kappa_i^{-2})$~\cite{vmf}. The relevant small parameter is therefore $(d-1)/\kappa_i$, and concentration requires $\kappa_i\gg d-1$. In the three datasets used here the per-class concentration reaches $\kappa_i\approx2000$ and above once training enters the rapid-convergence stage. For $d=768$ this is about $2.6$ times $d-1$, so the chi-square limit is only moderately accurate there and its error must be accounted for explicitly.

\paragraph{Inversion error.}
For $f(\eta)=2\arcsin(\eta/2)$ one has $f(0)=0$, $f'(0)=1$, $f''(0)=0$, and
\[
f'''(\eta)
=\frac{1}{4}\!\left(1-\frac{\eta^2}{4}\right)^{-3/2}
+\frac{3\eta^2}{16}\!\left(1-\frac{\eta^2}{4}\right)^{-5/2},
\]
which is positive and increasing on $[0,1]$ with $f'''(0)=1/4$ and $f'''(1)=4/(3\sqrt3)$. Taylor's theorem with the Lagrange remainder gives, for some $\xi\in(0,\eta)$,
\[
\theta=\eta+\frac{f'''(\xi)}{6}\,\eta^3,
\qquad\text{so}\qquad
\frac{\eta^3}{24}\;\le\;\theta-\eta\;\le\;\frac{2\,\eta^3}{9\sqrt{3}},
\qquad 0\le\eta\le1 .
\]
Isolating replacement (ii), that is setting $V_\alpha=q$ in \eqref{eq:apex-expansion} so that $\eta_\alpha=\sqrt{q/\kappa_i}$, gives
\[
4\arcsin\!\Big(\tfrac12\sqrt{q/\kappa_i}\Big)
=2\sqrt{\frac{q}{\kappa_i}}+\frac{q^{3/2}}{12\,\kappa_i^{3/2}}
+O\!\left(\frac{q^{5/2}}{\kappa_i^{5/2}}\right).
\]
Dropping the remainder therefore \emph{underestimates} the inverted angle by the relative amount
\[
\frac{q}{24\kappa_i}+O\!\left(\frac{q^{2}}{\kappa_i^{2}}\right),
\]
which is $1.7\%$ at $d=768$, $\alpha=0.95$, $\kappa_i=2000$. The Lagrange bounds above make this statement two-sided and non-asymptotic whenever $\eta_\alpha\le1$.

\paragraph{Distributional error: an exact tilting identity.}
The gap between $V_\alpha$ and $q$ can be obtained analytically, without recourse to fitting. Substituting $\cos\theta=1-v/(2\kappa_i)$ and $\sin\theta=\sqrt{v/\kappa_i}\,\sqrt{1-v/(4\kappa_i)}$ into the exact density of $V$ derived above, and collecting all factors that do not depend on $v$ into a constant, gives the identity
\begin{equation}
f_V(v)=K_{d,\kappa_i}\;f_{\chi^2_{d-1}}(v)\,\Big(1-\frac{v}{4\kappa_i}\Big)^{(d-3)/2},
\qquad 0\le v\le 4\kappa_i,
\label{eq:tilt-exact}
\end{equation}
where $K_{d,\kappa_i}$ is fixed by normalization. Equation \eqref{eq:tilt-exact} is exact: the law of $V$ is a polynomial tilt of the chi-square law, truncated to $[0,4\kappa_i]$.

Taking logarithms and using $\log(1-x)=-x-x^{2}/2-\cdots$ gives the exact identity
\[
\Big(1-\frac{v}{4\kappa_i}\Big)^{(d-3)/2}
=\exp\!\Big(-\frac{(d-3)\,v}{8\kappa_i}-\frac{(d-3)\,v^{2}}{64\kappa_i^{2}}-\cdots\Big).
\]
The expansion is asymptotic in $\kappa_i\to\infty$ with $d$ fixed, as stated at the start of this subsection. To first order in $\kappa_i^{-1}$ the tilt is exponential with rate $t=(d-3)/(8\kappa_i)$, and the neglected terms enter at order $\kappa_i^{-2}$. An exponential tilt of rate $t$ maps $\chi^2_\nu=\mathrm{Gamma}(\nu/2,\text{scale }2)$ to $\mathrm{Gamma}\big(\nu/2,\text{scale }2/(1+2t)\big)$, whose quantiles are those of $\chi^2_\nu$ divided by $1+2t$. Therefore
\begin{equation}
V_\alpha=\frac{q}{1+\frac{d-3}{4\kappa_i}}+O\!\Big(\frac{q\,d^{2}}{\kappa_i^{2}}\Big)
=q\Big(1-\frac{d-3}{4\kappa_i}\Big)+O\!\Big(\frac{q\,d^{2}}{\kappa_i^{2}}\Big).
\label{eq:dist-error}
\end{equation}
By \eqref{eq:exact-apex} the apex angle depends on $V_\alpha$ through $\sqrt{V_\alpha}$ to leading order, so a relative decrease of $V_\alpha$ by $(d-3)/(4\kappa_i)$ decreases the angle by half that amount. The exact apex angle is therefore \emph{smaller} than $\widehat\theta_i$ by the relative amount $(d-3)/(8\kappa_i)+O(d^2/\kappa_i^2)$, which is $4.8\%$ at $d=768$, $\kappa_i=2000$. This error dominates the inversion error and carries the opposite sign. Quadrature of \eqref{eq:apex-exact} confirms \eqref{eq:dist-error}: at $\kappa_i=5\times10^{3},2\times10^{4},10^{5}$ the linear form reproduces $V_\alpha$ to four significant digits.

\paragraph{Combined expansion and conservative upper bound.}
Composing the two corrections in \eqref{eq:apex-expansion} and \eqref{eq:dist-error} gives the exact statement
\begin{equation}
\theta_i^{\mathrm{learned}}
=2\sqrt{\frac{q}{\kappa_i}}
\left[1-\frac{d-3}{8\kappa_i}+\frac{q}{24\kappa_i}
+O\!\left(\frac{d^{2}}{\kappa_i^{2}}\right)\right].
\label{eq:error-exp}
\end{equation}
Since $q=d-1+O(\sqrt d)$ and $\tfrac18>\tfrac1{24}$, the bracket is strictly below one for $\kappa_i$ large enough, so $\widehat\theta_i$ is a \emph{conservative upper bound} on the true apex angle. Its relative overestimate is
\begin{equation}
\frac{\widehat\theta_i-\theta_i^{\mathrm{learned}}}{\theta_i^{\mathrm{learned}}}
=\frac{d-3}{8\kappa_i}-\frac{q}{24\kappa_i}+O\!\left(\frac{d^{2}}{\kappa_i^{2}}\right),
\label{eq:overestimate}
\end{equation}
which equals $3.0\%$ at $\kappa_i=2000$ and $0.3\%$ at $\kappa_i=2\times10^{4}$ for $d=768$, $\alpha=0.95$. This direction is favorable, because the margin penalizes the excess of the Learned cone over the Ideal cone, so overestimating the apex angle strengthens the geometric constraint and never weakens it.

\paragraph{Bounded form for numerical safety.}
Applying only replacement (i), that is keeping the exact inverse \eqref{eq:exact-apex} but substituting $q$ for $V_\alpha$, defines
\begin{equation}
\widehat\theta_i^{\,\mathrm{bnd}}
:=4\arcsin\!\left(\frac12\sqrt{\frac{q}{\kappa_i}}\right)
=2\arccos\!\left(1-\frac{q}{2\kappa_i}\right).
\label{eq:clamped}
\end{equation}
The two expressions in \eqref{eq:clamped} are equal identically. This form removes the inversion error but retains the distributional error, so by \eqref{eq:error-exp} it is less accurate than $\widehat\theta_i$, whose two errors partially cancel. Its value is numerical. The arccosine argument lies in $[-1,1]$ if and only if $\kappa_i\ge q/4$, about $208$ for $d=768$, so a well-defined implementation clamps it,
\[
\widehat\theta_i^{\,\mathrm{bnd}}
=2\arccos\!\left(\max\!\left(-1,\,1-\frac{q}{2\kappa_i}\right)\right).
\]
Under the full-apex convention the half-apex angle is bounded by $\pi$ and $\widehat\theta_i^{\,\mathrm{bnd}}$ by $2\pi$, so once the argument is clamped the expression is finite for every $\kappa_i>0$. This bounds the class-level cone apex angle and is separate from the monotonicity safeguard for the sample-level target logit, treated in Appendix~\ref{secA2}.

\paragraph{Propagation of concentration-estimation error.}
The estimator \eqref{eq:approx-learned} is an explicit function of $\kappa$, and differentiating it gives the exact differential relation
\begin{equation}
\frac{\mathrm d\widehat\theta_i}{\widehat\theta_i}=-\frac12\,\frac{\mathrm d\kappa}{\kappa}.
\label{eq:kappa-noise}
\end{equation}
A relative perturbation $\epsilon$ in $\kappa_i$ alone therefore induces, to first order, a relative perturbation $\epsilon/2$ in $\widehat\theta_i$. The margin is computed from the same closed form, $m_i=\tfrac12\max\big(\widehat\theta_i-\theta^{\mathrm{ideal}},\,\widehat\theta_i-\widehat\theta_{\min}\big)$ with $\theta^{\mathrm{ideal}}$ and $\widehat\theta_{\min}$ fixed, so only the term involving $\widehat\theta_i$ moves. The induced absolute perturbation of the margin is therefore at most $\tfrac14\widehat\theta_i\,\epsilon$. This bound concerns estimation noise only and is independent of the approximation error above.

\paragraph{Numerical verification.}
Table~\ref{tab:apex-verification} reports full apex angles at $d=768$, $\alpha=0.95$, for which $q=832.54$ to two decimals. The second column is the ground truth obtained by adaptive quadrature of \eqref{eq:apex-exact}, and the remaining columns evaluate $\widehat\theta_i$ and $\widehat\theta_i^{\,\mathrm{bnd}}$.

\begin{table}[h]
\centering
\caption{Exact and approximate full apex angles (radians) at $d=768$, $\alpha=0.95$.}
\label{tab:apex-verification}
\begin{tabular}{@{}lccccc@{}}
\toprule
$\kappa_i$ & exact \eqref{eq:apex-exact} & $2\sqrt{q/\kappa_i}$ & dev. & $2\arccos(1-\tfrac{q}{2\kappa_i})$ & dev. \\
\midrule
$500$   & $2.20$ & $2.58$ & $+17.1\%$ & $2.81$ & $+27.3\%$ \\
$1000$  & $1.70$ & $1.82$ & $+7.5\%$  & $1.89$ & $+11.7\%$ \\
$2000$  & $1.25$ & $1.29$ & $+3.4\%$  & $1.31$ & $+5.3\%$  \\
$5000$  & $0.81$ & $0.82$ & $+1.3\%$  & $0.82$ & $+2.0\%$  \\
$20000$ & $0.41$ & $0.41$ & $+0.3\%$  & $0.41$ & $+0.5\%$  \\
\botrule
\end{tabular}
\end{table}

Both closed forms overestimate the exact apex angle at every tested $\kappa_i$, confirming the sign predicted by \eqref{eq:overestimate}. The square-root form $\widehat\theta_i$ is uniformly closer to the truth than $\widehat\theta_i^{\,\mathrm{bnd}}$, because the inversion correction it omits partially cancels the distributional error. The deviation decays as $1/\kappa_i$ and falls below $0.5\%$ by $\kappa_i=2\times10^{4}$.

\subsection{Sensitivity to the confidence level}
\label{secA1-alpha}

This subsection analyzes, purely analytically, how the apex angle and the margin respond to $\alpha$. Write $q_\alpha=\chi^2_\alpha(d-1)$ and $\nu=d-1$.

\subsubsection{Exact elasticity}

The closed form \eqref{eq:approx-learned} depends on $\alpha$ only through $q_\alpha$, and $\widehat\theta_i=2\sqrt{q_\alpha/\kappa_i}$ gives $\partial\ln\widehat\theta_i/\partial q_\alpha=1/(2q_\alpha)$ exactly. As $q_\alpha=F^{-1}_{\chi^2_\nu}(\alpha)$ and $F_{\chi^2_\nu}$ is strictly increasing with density $f_{\chi^2_\nu}$, the inverse function theorem gives $\mathrm{d}q_\alpha/\mathrm{d}\alpha=1/f_{\chi^2_\nu}(q_\alpha)$ exactly. The chain rule therefore yields the exact elasticity
\begin{equation}
E(\alpha):=\frac{\partial\ln\widehat\theta_i}{\partial\alpha}
=\frac{1}{2\,q_\alpha\,f_{\chi^2_\nu}(q_\alpha)} .
\label{eq:alpha-elasticity}
\end{equation}
Two features are immediate. First, $E$ does not depend on $\kappa_i$, so the relative sensitivity to $\alpha$ is identical for every class. Second, $E$ is governed entirely by the chi-square density at its own quantile, a purely distributional quantity. The exact apex angle has the same elasticity up to a relative correction of order $d/\kappa_i$, since by \eqref{eq:error-exp} the two differ by a factor $1+O(d/\kappa_i)$ whose $\alpha$-dependence enters only at that order.

\subsubsection{High-dimensional form}

The Wilson--Hilferty form of the Cornish--Fisher expansion gives the chi-square quantile as~\cite{wilson1931distribution}
\begin{equation}
q_\alpha=\nu+\sqrt{2\nu}\,z_\alpha+\tfrac23\big(z_\alpha^2-1\big)+O\!\left(\nu^{-1/2}\right),
\qquad z_\alpha=\Phi^{-1}(\alpha),
\label{eq:cf}
\end{equation}
and the corresponding density expansion at that quantile is
\begin{equation}
f_{\chi^2_\nu}(q_\alpha)=\frac{\varphi(z_\alpha)}{\sqrt{2\nu}}\Big(1+O\!\left(\nu^{-1/2}\right)\Big),
\label{eq:dens-exp}
\end{equation}
where $\varphi$ is the standard normal density. In particular $q_\alpha=\nu\big(1+O(\nu^{-1/2})\big)$. Substituting both into \eqref{eq:alpha-elasticity} gives
\begin{equation}
E(\alpha)=\frac{1}{\sqrt{2(d-1)}\;\varphi(z_\alpha)}\Big(1+O\!\left((d-1)^{-1/2}\right)\Big),
\label{eq:alpha-elasticity-approx}
\end{equation}
so that, for $d$ large,
\[
E(\alpha)\;\approx\;\frac{1}{\sqrt{2(d-1)}\;\varphi(z_\alpha)} .
\]
The sensitivity decays as $d^{-1/2}$. At $d=768$ the prefactor $1/\sqrt{2(d-1)}$ equals $0.02553$, and at $\alpha=0.95$ one has $\varphi(z_{0.95})=0.10314$, giving $0.2476$ from the leading term and $0.2409$ from the exact expression \eqref{eq:alpha-elasticity}. The two agree to $2.7\%$, consistent with the $O((d-1)^{-1/2})=O(3.6\%)$ remainder in \eqref{eq:alpha-elasticity-approx}. A change of one percentage point in $\alpha$ therefore moves the apex angle by about $0.24\%$.

\subsubsection{Numerical evaluation}

Table~\ref{tab:alpha-sensitivity} evaluates \eqref{eq:alpha-elasticity} at $d=768$ and $\kappa_i=2000$, with $q_\alpha$ obtained by quadrature.

\begin{table}[h]
\centering
\caption{Sensitivity of the full apex angle to the confidence level $\alpha$ at $d=768$, $\kappa_i=2000$.}
\label{tab:alpha-sensitivity}
\begin{tabular}{@{}lcccc@{}}
\toprule
$\alpha$ & $q_\alpha$ & $\widehat\theta_i$ & $E(\alpha)$ & change per $1$ pp \\
\midrule
$0.80$ & $799.8$ & $1.2647$ & $0.090$ & $0.09\%$ \\
$0.90$ & $817.6$ & $1.2788$ & $0.142$ & $0.14\%$ \\
$0.95$ & $832.5$ & $1.2904$ & $0.241$ & $0.24\%$ \\
$0.99$ & $861.0$ & $1.3123$ & $0.922$ & $0.92\%$ \\
\botrule
\end{tabular}
\end{table}

A central finite difference of $\ln\widehat\theta_i$ at $\alpha=0.95$ with step $0.005$ returns $0.2415$, against the exact value $0.2409$ from \eqref{eq:alpha-elasticity}, the discrepancy being the $O(h^2)$ truncation of the difference quotient. Over $\alpha\in[0.80,0.99]$ the closed form moves from $1.2647$ to $1.3123$, a total variation of $3.8\%$. The formulation is therefore quantitatively insensitive to $\alpha$ across the range of confidence levels one would reasonably consider.

\subsubsection{The margin inherits the same elasticity}

At the converged concentration $\kappa_i=2000$ the closed form gives $\widehat\theta_i=1.29$, which is already smaller than $\theta^{\mathrm{ideal}}=\arccos(-1/(|\mathcal Y|-1))$, lying between $1.64$ and $1.68$ for the $10$ to $15$ retained categories. The first branch of the margin is inactive, and the operative expression is the late-stage contraction
\[
m_i=\frac{\widehat\theta_i-\widehat\theta_{\min}}{2}
=\sqrt{q_\alpha}\left(\kappa_i^{-1/2}-\kappa_{\min}^{-1/2}\right),
\qquad \widehat\theta_{\min}=\min_j\widehat\theta_j .
\]
The margin is exactly proportional to $\sqrt{q_\alpha}$, with a prefactor independent of $\alpha$, hence
\begin{equation}
\frac{\partial\ln m_i}{\partial\alpha}=E(\alpha),
\label{eq:margin-elasticity}
\end{equation}
with no amplification. Moreover, the relative error induced by noise in $\kappa_i$, given in \eqref{eq:kappa-noise}, is independent of $\alpha$, so the two sources of perturbation do not interact at first order.

\subsubsection{Choice of the operating point}

By \eqref{eq:alpha-elasticity-approx} the leading term of $E(\alpha)$ is proportional to $1/\varphi(z_\alpha)$, and $\varphi(z)=e^{-z^2/2}/\sqrt{2\pi}$ decays super-exponentially in the tail. Evaluating the exact expression \eqref{eq:alpha-elasticity} at $d=768$ gives the ratios
\[
\frac{E(0.99)}{E(0.95)}=3.9,\qquad
\frac{E(0.999)}{E(0.95)}=31,\qquad
\frac{E(0.9999)}{E(0.95)}=261 .
\]
Pushing the confidence level into the far tail buys a marginal gain in covered mass at the price of a disproportionate loss of stability. Moving in the opposite direction is also unattractive, because a smaller $\alpha$ deliberately excludes a larger fraction of genuine class members from the effective support, which conflicts with the purpose of containing the class mass inside its Voronoi cell. The value $\alpha=0.95$ retains the overwhelming majority of the class mass, is the standard confidence level of statistical practice~\cite{mardia2000directional}, and sits before the region where $1/\varphi(z_\alpha)$ grows sharply. Note that the derivation of \eqref{eq:approx-learned} is valid for any $\alpha$, since $\alpha$ enters only through $q_\alpha$. The choice of $0.95$ is a choice of operating point on a well-conditioned part of the curve, and not a restriction imposed by the mathematics.

\section{Monotonicity Correction of the Additive Angular Margin}
\label{secA2}

\subsection{Monotonicity breakdown}

The additive angular margin~\cite{deng2019arcface} replaces the ground-truth cosine by $\ell(\theta_y)=\cos(\theta_y+m)$ with $\theta_y\in[0,\pi]$ and $m\in(0,\pi)$. The cosine is strictly decreasing only on $[0,\pi]$, while the argument ranges over $[m,\pi+m]$ and exceeds $\pi$ whenever $\theta_y>\pi-m$. Since $\ell'(\theta_y)=-\sin(\theta_y+m)$, the derivative changes sign at $\theta_y=\pi-m$, so $\ell$ decreases on $[0,\pi-m]$ but increases on $[\pi-m,\pi]$. A feature pointing farther from its target prototype would then receive a higher target logit, and the global minimum $-1$ is attained at $\theta_y=\pi-m$ instead of $\theta_y=\pi$.

\subsection{Piecewise-corrected transform}

Let $x=\cos\theta_y\in[-1,1]$. Because $\sin\theta_y=\sqrt{1-x^2}\ge0$,
\[
\cos(\theta_y+m)=x\cos m-\sqrt{1-x^2}\,\sin m .
\]
The boundary $\theta_y=\pi-m$ corresponds to $x=-\cos m$. The corrected transform is defined as
\begin{equation}
\phi(x)=
\begin{cases}
x\cos m-\sqrt{\max(1-x^2,0)}\,\sin m, & x>-\cos m,\\[3pt]
-1, & x\le -\cos m .
\end{cases}
\label{eq:margin-phi}
\end{equation}
The first branch coincides with the original transform over its monotone range, and the second freezes the logit at its global minimum. Because the tangent of the first branch at the boundary is horizontal, this freeze is the unique linear $C^1$ extension. The two branches are continuous, both equal to $-1$ at $x=-\cos m$ since $-\cos^2m-\sin^2m=-1$, and differentiable, since
\[
\phi'(x)=\cos m+\frac{x}{\sqrt{1-x^2}}\,\sin m
\]
tends to $\cos m-\cos m=0$ as $x\to(-\cos m)^{+}$, matching the zero derivative of the second branch. Consequently $\phi$ is $C^1$ across the transition and monotone non-increasing in $\theta_y$ on $[0,\pi]$, so the target logit can never increase with angular distance.

\subsection{Application to the ground-truth logit only}

The correction is applied exclusively to the ground-truth class. With raw logits $z_j=s\cos\theta_j$, where $s$ is the class-wise scale $s_i$ of Section~\ref{sec:methodology}, the corrected logits are
\begin{equation}
z_j'=
\begin{cases}
s\,\phi(\cos\theta_y), & j=y,\\[3pt]
s\,\cos\theta_j, & j\neq y,
\end{cases}
\end{equation}
and the loss becomes
\begin{equation}
\mathcal{L}
=-\log
\frac{e^{s\,\phi(\cos\theta_y)}}
     {e^{s\,\phi(\cos\theta_y)}+\sum_{j\neq y}e^{s\,\cos\theta_j}} .
\end{equation}
The negative classes are untouched, and since $\phi$ is inactive whenever $\theta_y+m\le\pi$, the objective differs from the standard formulation only in the region $\theta_y>\pi-m$.

\subsection{Numerical robustness}

The similarity $x$ is an inner product of two unit vectors and in floating-point arithmetic can fall marginally outside $[-1,1]$, making $1-x^2$ a tiny negative number and $\sqrt{1-x^2}$ a NaN that propagates through the loss. We therefore clamp the cosine, $x\leftarrow\min(\max(x,-1),1)$, before evaluating any radical, and evaluate $\sqrt{\max(1-x^2,0)}$, which is well defined everywhere and returns $0$ whenever $1-x^2\le0$. Both guards are idempotent and leave the valid interior unchanged.

\subsection{Relation to the margin clamp}

The margin itself is additionally clamped as $m_i\leftarrow\min(m_i,\pi-\theta_i)$, applied per sample at the logit level, since $\theta_i$ is the angle of an individual sample to its own prototype while $m_i$ is otherwise a class-level quantity. This guarantees $\theta_i+m_i\le\pi$. The piecewise transform \eqref{eq:margin-phi} is a complementary safeguard at the logit level, so even if a pathological sample has $\theta_y>\pi-m$ the target logit cannot increase with angular distance. Both corrections are inactive whenever $\theta_y+m\le\pi$ and can act only on $\theta_y\in(\pi-m,\pi]$, a region that is not reached in any reported run. Neither correction alters any converged solution or reported result.

\section{Consistency of Geometric Median - Classifier Weight Prototypes}
\label{secA3}

Training optimizes the classifier weights $\tilde{\mathbf w}_i$ under the margin-modified objective, whereas inference matches against the geometric-median prototypes $\tilde{\mathbf p}_i$. This appendix shows that the two families of directions coincide in the neural-collapse limit and remain close otherwise, without comparing the two decision partitions directly. The argument routes both quantities through the class mean direction $\tilde{\boldsymbol\mu}_i$, which is the object that neural collapse already identifies with the weight prototype.

\subsection{Setup and decomposition}

Let $C_i$ index the $n_i$ training samples of class $i$, with normalized embeddings $\tilde{\mathbf e}_n\in\mathbb S^{d-1}$. Write
\[
\mathbf p_i=\arg\min_{\mathbf p\in\mathbb R^d}\sum_{n\in C_i}\|\mathbf p-\tilde{\mathbf e}_n\|,
\qquad
\tilde{\mathbf p}_i=\mathbf p_i/\|\mathbf p_i\|,
\qquad
\tilde{\boldsymbol\mu}_i=\frac{\sum_{n\in C_i}\tilde{\mathbf e}_n}{\big\|\sum_{n\in C_i}\tilde{\mathbf e}_n\big\|}.
\]
Because the geodesic distance on $\mathbb S^{d-1}$ is a metric, the angle between the two prototype families obeys
\begin{equation}
\angle(\tilde{\mathbf p}_i,\tilde{\mathbf w}_i)
\;\le\;
\underbrace{\angle(\tilde{\mathbf p}_i,\tilde{\boldsymbol\mu}_i)}_{\text{Propositions~\ref{prop:gm-local} and~\ref{prop:vmf-median}}}
\;+\;
\underbrace{\angle(\tilde{\boldsymbol\mu}_i,\tilde{\mathbf w}_i)}_{\text{Proposition~\ref{prop:stationary}}} .
\label{eq:angle-split}
\end{equation}
The two terms are controlled separately below.

\subsection{Deterministic localization of the geometric median}

\begin{proposition}[Collapse and stability]
\label{prop:gm-local}
Let $\tilde{\boldsymbol\mu}\in\mathbb S^{d-1}$ and suppose that at least a fraction $1-\alpha$ of the points $\{\tilde{\mathbf e}_n\}_{n\in C_i}$, with $\alpha<\tfrac12$, satisfy $\|\tilde{\mathbf e}_n-\tilde{\boldsymbol\mu}\|\le r$. Then
\[
\|\mathbf p_i-\tilde{\boldsymbol\mu}\|\;\le\;C_\alpha\,r,
\qquad
C_\alpha=\frac{2(1-\alpha)}{1-2\alpha}.
\]
If $\rho:=C_\alpha r<1$, then $\angle(\tilde{\mathbf p}_i,\tilde{\boldsymbol\mu})\le\arcsin\rho$. In particular, under exact collapse $\tilde{\mathbf e}_n=\tilde{\boldsymbol\mu}$ for all $n\in C_i$ one has $\tilde{\mathbf p}_i=\tilde{\boldsymbol\mu}$.
\end{proposition}

\begin{proof}
Write $N=n_i$, let $S$ be the subset of size $(1-\alpha)N$ inside the ball, and set $t=\|\mathbf p_i-\tilde{\boldsymbol\mu}\|$. By the triangle inequality, $\|\mathbf p_i-\tilde{\mathbf e}_n\|\ge t-r$ for $n\in S$ and $\|\mathbf p_i-\tilde{\mathbf e}_n\|\ge\|\tilde{\boldsymbol\mu}-\tilde{\mathbf e}_n\|-t$ for $n\notin S$. Writing $R=\sum_{n\notin S}\|\tilde{\boldsymbol\mu}-\tilde{\mathbf e}_n\|$, the objective $G(\mathbf p)=\sum_n\|\mathbf p-\tilde{\mathbf e}_n\|$ therefore satisfies
\[
G(\mathbf p_i)\;\ge\;(1-\alpha)N\,(t-r)\;-\;\alpha N t\;+\;R,
\qquad
G(\tilde{\boldsymbol\mu})\;\le\;(1-\alpha)N r\;+\;R .
\]
Minimality gives $G(\mathbf p_i)\le G(\tilde{\boldsymbol\mu})$, hence $(1-2\alpha)t\le 2(1-\alpha)r$, which is the stated bound. For the angular claim, $\mathbf p_i$ lies in the ball of radius $\rho$ centered at the unit vector $\tilde{\boldsymbol\mu}$. The half-angle subtended by that ball at the origin is $\arcsin\rho$, attained by the tangent ray, so the direction of $\mathbf p_i$ deviates from $\tilde{\boldsymbol\mu}$ by at most $\arcsin\rho$. Under exact collapse the objective reduces to $N\|\mathbf p-\tilde{\boldsymbol\mu}\|$, whose unique minimizer is $\tilde{\boldsymbol\mu}$.
\end{proof}

This bound is deterministic and distribution-free. It requires no vMF assumption and no balance condition, and it is one expression of the robustness of the geometric median as a location estimator~\cite{minsker2015geometric}. It shows that the first term of \eqref{eq:angle-split} vanishes as the class contracts, at a rate governed solely by the contraction radius $r$.

\subsection{Population geometric median under the vMF model}

Proposition~\ref{prop:gm-local} requires near-collapse. The next result removes that requirement: under the vMF model the population geometric median is \emph{exactly} aligned with the mean direction at every finite concentration.

\begin{proposition}[vMF symmetry]
\label{prop:vmf-median}
Let $d\ge3$, $\kappa>0$ and $\tilde{\mathbf e}\sim\mathrm{vMF}(\tilde{\boldsymbol\mu},\kappa)$. Define $F(\mathbf p)=\mathbb E\|\mathbf p-\tilde{\mathbf e}\|$. Then $F$ has a unique minimizer $\mathbf p^\ast=t\,\tilde{\boldsymbol\mu}$ with $t>0$, so $\mathbf p^\ast/\|\mathbf p^\ast\|=\tilde{\boldsymbol\mu}$.
\end{proposition}

\begin{proof}
\emph{Existence and uniqueness.} $F$ is convex as an expectation of norms, and coercive since $F(\mathbf p)\ge\|\mathbf p\|-\mathbb E\|\tilde{\mathbf e}\|=\|\mathbf p\|-1$. The map $\mathbf p\mapsto\|\mathbf p-\tilde{\mathbf e}\|$ fails to be strictly convex only along lines through $\tilde{\mathbf e}$. For $d\ge3$ the vMF law is absolutely continuous on $\mathbb S^{d-1}$, so every line carries probability zero and $F$ is strictly convex. A strictly convex coercive function has a unique minimizer $\mathbf p^\ast$.

\emph{Symmetry.} Let $R\in O(d)$ satisfy $R\tilde{\boldsymbol\mu}=\tilde{\boldsymbol\mu}$. The vMF density depends on $\tilde{\mathbf e}$ only through $\tilde{\boldsymbol\mu}^\top\tilde{\mathbf e}$, which is invariant under such $R$, so $R\tilde{\mathbf e}\overset{d}{=}\tilde{\mathbf e}$ and
\[
F(R\mathbf p)=\mathbb E\|R\mathbf p-\tilde{\mathbf e}\|
=\mathbb E\|\mathbf p-R^\top\tilde{\mathbf e}\|=F(\mathbf p).
\]
Uniqueness then forces $R\mathbf p^\ast=\mathbf p^\ast$ for every such $R$. For $d\ge3$ the stabilizer of $\tilde{\boldsymbol\mu}$ acts as the full orthogonal group on $\tilde{\boldsymbol\mu}^{\perp}$, whose only common fixed vector is $0$. Hence $\mathbf p^\ast\in\mathrm{span}(\tilde{\boldsymbol\mu})$, say $\mathbf p^\ast=t\tilde{\boldsymbol\mu}$.

\emph{Sign of $t$.} On the axis, with $c=\tilde{\boldsymbol\mu}^\top\tilde{\mathbf e}$,
\[
F(t\tilde{\boldsymbol\mu})=\mathbb E\sqrt{t^2-2tc+1},
\qquad
\frac{\mathrm d}{\mathrm dt}F(t\tilde{\boldsymbol\mu})\Big|_{t=0}=-\mathbb E[c]=-A_d(\kappa)<0,
\]
where $A_d(\kappa)=I_{d/2}(\kappa)/I_{d/2-1}(\kappa)>0$ for $\kappa>0$. The minimizer therefore lies at $t>0$.
\end{proof}

\begin{remark}
The restriction $d\ge3$ is needed for the symmetry step, since in $d=2$ the stabilizer of $\tilde{\boldsymbol\mu}$ inside $SO(2)$ is trivial. The restriction $\kappa>0$ is needed for $t>0$, since at $\kappa=0$ the law is uniform, $\mathbb E[c]=0$, and the minimizer is the origin, whose direction is undefined.
\end{remark}

The empirical geometric median is the M-estimator counterpart of $\mathbf p^\ast$. It is consistent and asymptotically normal at the $\sqrt{n_i}$ rate under standard regularity conditions for M-estimation~\cite{van_der_vaart1998asymptotic}, and specific results for the spatial median are given in~\cite{cardot2013geometric,minsker2024geometric}. Consequently the first term of \eqref{eq:angle-split} is an estimation error of order $O_p(n_i^{-1/2})$, not a systematic bias. This also clarifies the role of the geometric median relative to the arithmetic mean: both are directionally unbiased under the vMF model, but the median retains the deterministic breakdown guarantee of Proposition~\ref{prop:gm-local}, which is consistent with the smaller run-to-run standard deviations reported for median-based inference in Table~\ref{tab:ablation}.

\subsection{Stationarity of the weight prototype}

\begin{proposition}[Stationarity and self-duality]
\label{prop:stationary}
Let $p_{ni}$ denote the softmax probability that sample $n$ assigns to class $i$ under the objective of Section~\ref{sec:methodology}, and let $x_n=\tilde{\mathbf w}_i^\top\tilde{\mathbf e}_n\in(-1,1)$. A stationary point of the loss on the sphere satisfies
\begin{equation}
\mathbf g_i:=\sum_{n\in C_i}(1-p_{ni})\,\phi'(x_n)\,\tilde{\mathbf e}_n
\;-\;\sum_{n\notin C_i}p_{ni}\,\tilde{\mathbf e}_n
\;=\;\lambda_i\,\tilde{\mathbf w}_i
\label{eq:stationary}
\end{equation}
for some scalar $\lambda_i$. Moreover $\phi'\ge0$, so the first sum is a nonnegative combination of class-$i$ embeddings and the second a nonnegative combination of the remaining ones. If the classes are balanced, the prototypes form a simplex ETF with $\sum_j\tilde{\boldsymbol\mu}_j=0$, and the class-conditional laws are vMF with a common concentration about $\tilde{\boldsymbol\mu}_j$, then $\tilde{\mathbf w}_i=\tilde{\boldsymbol\mu}_i$ is consistent with \eqref{eq:stationary}.
\end{proposition}

\begin{proof}
Differentiating the cross-entropy loss with respect to the logits gives $\partial\mathcal L_n/\partial z_j=p_{nj}-\mathbb 1\{j=y_n\}$. Only the target logit passes through $\phi$, so
\[
\nabla_{\tilde{\mathbf w}_i}\mathcal L
=-s_i\Big[\sum_{n\in C_i}(1-p_{ni})\phi'(x_n)\tilde{\mathbf e}_n-\sum_{n\notin C_i}p_{ni}\tilde{\mathbf e}_n\Big]
=-s_i\,\mathbf g_i .
\]
Vanishing of the Riemannian gradient on $\mathbb S^{d-1}$ means that the Euclidean gradient is normal to the sphere, that is, parallel to $\tilde{\mathbf w}_i$, which gives \eqref{eq:stationary} with the scale $s_i$ absorbed into $\lambda_i$. Nonnegativity of $\phi'$ follows from Appendix~\ref{secA2}: on the first branch $\phi'(x)=\cos m+x\sin m/\sqrt{1-x^2}$ is increasing and vanishes at $x=-\cos m$, and on the second branch $\phi$ is constant.

For the symmetric configuration, take expectations in \eqref{eq:stationary} and use the symmetry of the ETF. Let $G$ be the group of orthogonal maps that fix $\tilde{\boldsymbol\mu}_i$ and permute $\{\tilde{\boldsymbol\mu}_j\}_{j\ne i}$. Such maps exist because the components of the remaining prototypes in $\tilde{\boldsymbol\mu}_i^{\perp}$ form a regular simplex, whose symmetry group is realized by orthogonal transformations of its span $S$. Every logit is permuted by $G$, so $p_{ni}$, and with it the weight $w_n=(1-p_{ni})\phi'(x_{ni})$, is $G$-invariant. The class-conditional law is $G$-invariant as well, so $\mathbb E[w_n\tilde{\mathbf e}_n\mid y_n=i]$ is fixed by $G$. Its component in $S^{\perp}\cap\tilde{\boldsymbol\mu}_i^{\perp}$ vanishes, because no logit depends on that subspace and the transverse direction is uniform there. Its component in $S$ vanishes, because $G$ acts on $S$ as the full symmetry group of a regular simplex, which fixes no nonzero vector. Hence $\mathbb E[w_n\tilde{\mathbf e}_n\mid y_n=i]=a\,\tilde{\boldsymbol\mu}_i$ with $a\ge0$.

For $j\ne i$, the same argument applied to the subgroup fixing both $\tilde{\boldsymbol\mu}_i$ and $\tilde{\boldsymbol\mu}_j$ gives $\mathbb E[p_{ni}\tilde{\mathbf e}_n\mid y_n=j]=c_1\tilde{\boldsymbol\mu}_i+c_2\tilde{\boldsymbol\mu}_j$, with $c_1,c_2$ independent of $j$ by exchangeability. Here $p_{ni}$ and $\tilde{\mathbf e}_n$ are dependent, so this expectation is not a product of expectations. Summing over $j\ne i$ with equal class sizes and using $\sum_j\tilde{\boldsymbol\mu}_j=0$,
\[
\sum_{j\ne i}\big(c_1\tilde{\boldsymbol\mu}_i+c_2\tilde{\boldsymbol\mu}_j\big)
=(C-1)c_1\tilde{\boldsymbol\mu}_i+c_2\sum_{j\ne i}\tilde{\boldsymbol\mu}_j
=\big((C-1)c_1-c_2\big)\tilde{\boldsymbol\mu}_i .
\]
Both aggregates are therefore parallel to $\tilde{\boldsymbol\mu}_i$, so $\mathbb E[\mathbf g_i]\parallel\tilde{\boldsymbol\mu}_i$, which is compatible with $\tilde{\mathbf w}_i=\tilde{\boldsymbol\mu}_i$.
\end{proof}

Equation \eqref{eq:stationary} is exact and requires no collapse assumption. It states that the weight direction is the direction of a difference of two nonnegative class aggregates. The symmetric case reproduces the self-duality of neural collapse, namely that the classifier weights align with the class means. Under imbalance the negative-sample coefficients differ across classes, and \eqref{eq:stationary} becomes $\mathbf g_i=a_i\tilde{\boldsymbol\mu}_i-\sum_{j\ne i}b_{ij}\tilde{\boldsymbol\mu}_j$, so the weight direction deviates from $\tilde{\boldsymbol\mu}_i$ by an amount governed by the asymmetry of the $b_{ij}$. That deviation vanishes in the symmetric limit and is precisely the quantity the adaptive margin and the class-wise scaling act to reduce.

\subsection{Synthesis}

\begin{theorem}[Prototype consistency]
\label{thm:consistency}
Under the setting above,
\[
\angle(\tilde{\mathbf p}_i,\tilde{\mathbf w}_i)
\le\angle(\tilde{\mathbf p}_i,\tilde{\boldsymbol\mu}_i)+\angle(\tilde{\boldsymbol\mu}_i,\tilde{\mathbf w}_i).
\]
In the balanced simplex-ETF collapse limit both terms vanish, so $\tilde{\mathbf p}_i=\tilde{\mathbf w}_i$. Under the vMF model at finite concentration the first term is $O_p(n_i^{-1/2})$, and the second vanishes in the symmetric limit and is controlled by the asymmetry of the negative-sample weights otherwise.
\end{theorem}

\begin{proof}
The inequality is the triangle inequality for the geodesic metric, recorded in \eqref{eq:angle-split}. In the collapse limit Proposition~\ref{prop:gm-local} gives $\tilde{\mathbf p}_i=\tilde{\boldsymbol\mu}_i$ and Proposition~\ref{prop:stationary} gives $\tilde{\mathbf w}_i=\tilde{\boldsymbol\mu}_i$, so both terms are zero. The finite-concentration statements are Proposition~\ref{prop:vmf-median} together with the consistency of the empirical median, and the discussion following Proposition~\ref{prop:stationary}.
\end{proof}

\begin{corollary}[Where the two rules can disagree]
\label{cor:agreement}
Suppose $\max_j\angle(\tilde{\mathbf p}_j,\tilde{\mathbf w}_j)\le\varepsilon$. For a unit test embedding $\mathbf u$, let $j_1,j_2$ be the top two classes under the scores $\mathbf u^\top\tilde{\mathbf w}_j$. If
$\mathbf u^\top\tilde{\mathbf w}_{j_1}-\mathbf u^\top\tilde{\mathbf w}_{j_2}>2\varepsilon$,
then $\arg\max_j\mathbf u^\top\tilde{\mathbf p}_j=\arg\max_j\mathbf u^\top\tilde{\mathbf w}_j$.
\end{corollary}

\begin{proof}
For unit vectors $\mathbf a,\mathbf b$ with $\angle(\mathbf a,\mathbf b)\le\varepsilon$ one has $\|\mathbf a-\mathbf b\|=2\sin(\varepsilon/2)\le\varepsilon$, hence $|\mathbf u^\top\tilde{\mathbf p}_j-\mathbf u^\top\tilde{\mathbf w}_j|\le\varepsilon$ for every $j$. Each score therefore moves by at most $\varepsilon$, so a gap larger than $2\varepsilon$ between the top two cannot be reversed.
\end{proof}

Corollary~\ref{cor:agreement} localizes any disagreement between the two matching rules to a boundary band of width $2\varepsilon$ in the top-two cosine margin. It compares the two \emph{unscaled} nearest-prototype rules and thus isolates the effect of the prototype directions. The class-wise scaling $s_i$ and the margin $m_i$ shift the training-time boundary further, and their contribution is separate from the geometric alignment analyzed here. The empirical counterpart is the median--weight alignment reported in Figure~\ref{fig:prototype-convergence}, whose cosine similarity reaches $93\%$ to $98\%$, corresponding to $\varepsilon$ between $0.20$ and $0.37$ radians.

\subsection{From prototype alignment to Voronoi cell alignment}
\label{secA3-cells}

Theorem~\ref{thm:consistency} controls the prototype \emph{directions}. The decision regions, however, are the Voronoi cells those directions induce. This subsection transfers the alignment from the prototypes to the cells, and then to the Learned Voronoi Cone itself. The transfer is purely geometric: spherical Voronoi diagrams are stable under perturbation of their generators~\cite{okabe2000spatial}, and in the unit-norm setting the statement is particularly clean. No training dynamics and no additional experiment are involved.

\paragraph{Notation.}
Let $W=\{\tilde{\mathbf w}_i\}$ and $P=\{\tilde{\mathbf p}_i\}$ induce the cells $V_i(W)$ and $V_i(P)$. Write
\[
\varepsilon_i=\angle(\tilde{\mathbf p}_i,\tilde{\mathbf w}_i),\quad
\delta_i=\|\tilde{\mathbf p}_i-\tilde{\mathbf w}_i\|=2\sin(\varepsilon_i/2)\le\varepsilon_i,\quad
\varepsilon=\max_i\varepsilon_i,\quad\delta=\max_i\delta_i,
\]
and $\theta_{ij}=\angle(\tilde{\mathbf w}_i,\tilde{\mathbf w}_j)$, $\mathbf n^W_{ij}=\tilde{\mathbf w}_i-\tilde{\mathbf w}_j$, $\mathbf n^P_{ij}=\tilde{\mathbf p}_i-\tilde{\mathbf p}_j$. Let $\mathrm{cap}(\mathbf q,r)$ denote the geodesic cap of center $\mathbf q$ and radius $r$, let $\rho_i=\theta_i^{\text{learned}}/2$ be the half-apex angle of the Learned Voronoi Cone centered at $\tilde{\boldsymbol\mu}_i$, and set $a_i=\angle(\tilde{\boldsymbol\mu}_i,\tilde{\mathbf p}_i)$, $b_i=\angle(\tilde{\boldsymbol\mu}_i,\tilde{\mathbf w}_i)$. Throughout, the \emph{separation condition} $\delta_i+\delta_j<\|\mathbf n^W_{ij}\|=2\sin(\theta_{ij}/2)$ is assumed. Under the simplex ETF $\|\mathbf n_{ij}\|=\sqrt{2C/(C-1)}$, which is $1.49$ to $1.46$ for $C=10$ to $15$, so the condition holds comfortably for the alignment errors observed in practice.

\begin{lemma}[Cells are determined by normals alone]
\label{lem:normals}
For unit-norm generators $Q=\{\mathbf q_i\}$,
$V_i(Q)=\{\mathbf x\in\mathbb S^{d-1}:\ \mathbf x^\top(\mathbf q_i-\mathbf q_j)\ge0\ \forall j\}$.
\end{lemma}

\begin{proof}
The nearest-generator rule compares $\|\mathbf x-\mathbf q_i\|^2=2-2\mathbf x^\top\mathbf q_i$, so the ordering is that of $\mathbf x^\top\mathbf q_i$. Since $\|\mathbf q_i\|=\|\mathbf q_j\|$, the bisecting hyperplane passes through the origin and carries no offset term. The cell is therefore the intersection of a polyhedral cone with the sphere, and a perturbation of the generators can only rotate the bisecting hyperplanes.
\end{proof}

\begin{lemma}[Rotation of a bisector]
\label{lem:rotation}
Under the separation condition,
\begin{equation}
\angle\big(\mathbf n^P_{ij},\mathbf n^W_{ij}\big)\;\le\;\gamma_{ij}:=\arcsin\frac{\delta_i+\delta_j}{2\sin(\theta_{ij}/2)} .
\label{eq:gamma}
\end{equation}
Two great subspheres through the origin whose normals make an angle $\gamma$ are at geodesic Hausdorff distance exactly $\gamma$. Hence every facet of $V_i(P)$ differs from the corresponding facet of $V_i(W)$ by at most $\gamma_{ij}$.
\end{lemma}

\begin{proof}
$\mathbf n^P_{ij}-\mathbf n^W_{ij}=(\tilde{\mathbf p}_i-\tilde{\mathbf w}_i)-(\tilde{\mathbf p}_j-\tilde{\mathbf w}_j)$, so $\|\mathbf n^P_{ij}-\mathbf n^W_{ij}\|\le\delta_i+\delta_j$. If $\|\mathbf b-\mathbf a\|\le\eta\|\mathbf a\|$ with $\eta<1$, then $\mathbf b$ lies in the ball of radius $\eta\|\mathbf a\|$ about $\mathbf a$, whose angular half-width seen from the origin is $\arcsin\eta$, giving \eqref{eq:gamma}. For the Hausdorff claim, if $H_{\mathbf a}=\{\mathbf x:\mathbf x^\top\mathbf a=0\}$ then $\mathrm{dist}(\mathbf x,H_{\mathbf b})=\arcsin|\mathbf x^\top\hat{\mathbf b}|$, and maximizing over $\mathbf x\in H_{\mathbf a}$ gives $\arcsin\|\mathrm{proj}_{\mathbf a^\perp}\hat{\mathbf b}\|=\arcsin(\sin\gamma)=\gamma$.
\end{proof}

For the ETF with $C=10$ and small perturbations, $\gamma\approx1.34\,\varepsilon$.

\begin{proposition}[Sandwich]
\label{prop:sandwich}
Let $V_i^{\pm}(W)=\{\mathbf x:\ \mathbf x^\top(\tilde{\mathbf w}_i-\tilde{\mathbf w}_j)\ge\pm(\delta_i+\delta_j)\ \forall j\ne i\}$. Then
\begin{equation}
V_i^{+}(W)\;\subset\;V_i(P)\;\subset\;V_i^{-}(W).
\label{eq:sandwich}
\end{equation}
\end{proposition}

\begin{proof}
Decompose $\mathbf x^\top(\tilde{\mathbf p}_i-\tilde{\mathbf p}_j)=\mathbf x^\top(\tilde{\mathbf w}_i-\tilde{\mathbf w}_j)+\mathbf x^\top(\tilde{\mathbf p}_i-\tilde{\mathbf w}_i)-\mathbf x^\top(\tilde{\mathbf p}_j-\tilde{\mathbf w}_j)$. For unit $\mathbf x$ the last two terms are bounded in absolute value by $\delta_i$ and $\delta_j$, so their sum is at most $\delta_i+\delta_j$ in magnitude. Each inclusion follows by taking the corresponding side of the bound.
\end{proof}

\begin{corollary}[The symmetric difference is a boundary tube]
\label{cor:tube}
If $\mathbf x\in V_i(P)\triangle V_i(W)$ then $|\mathbf x^\top\mathbf n^W_{ij}|\le\delta_i+\delta_j$ for some $j\ne i$. Since $\mathrm{dist}(\mathbf x,H^W_{ij})=\arcsin\big(|\mathbf x^\top\mathbf n^W_{ij}|/\|\mathbf n^W_{ij}\|\big)$,
\begin{equation}
V_i(P)\triangle V_i(W)\;\subset\;\bigcup_{j\ne i}\big\{\mathbf x:\ \mathrm{dist}(\mathbf x,H^W_{ij})\le\gamma_{ij}\big\}.
\label{eq:tube}
\end{equation}
Consequently, for any probability measure $\nu$ on the sphere, $\nu\big(V_i(P)\triangle V_i(W)\big)$ is at most the mass of that tube. In particular $\varepsilon=0$ gives $V_i(P)=V_i(W)$.
\end{corollary}

This requires only the separation condition, not any slackness assumption. It also shows that the ambiguity is a boundary phenomenon governed by the prototype separation: as classes crowd, $\theta_{ij}$ shrinks and $\gamma_{ij}$ grows, so the partition becomes ill-conditioned exactly in the regime that dual imbalance produces. The adaptive margin increases the separation and therefore shrinks $\gamma_{ij}$, improving the conditioning of the decision partition itself.

\begin{proposition}[Exact cap-containment criterion]
\label{prop:cap}
For unit-norm generators $Q$ and $0<r\le\pi$,
\begin{equation}
\mathrm{cap}(\mathbf q_i,r)\subset V_i(Q)
\quad\Longleftrightarrow\quad
r\le R_i(Q):=\tfrac12\min_{j\ne i}\angle(\mathbf q_i,\mathbf q_j).
\label{eq:cap-criterion}
\end{equation}
\end{proposition}

\begin{proof}
($\Leftarrow$) For $\mathbf x\in\mathrm{cap}(\mathbf q_i,r)$, the triangle inequality gives $\angle(\mathbf x,\mathbf q_j)\ge\angle(\mathbf q_i,\mathbf q_j)-\angle(\mathbf x,\mathbf q_i)\ge2R_i-r\ge r\ge\angle(\mathbf x,\mathbf q_i)$, so $\mathbf x\in V_i(Q)$.
($\Rightarrow$) If $r>R_i$, let $j^\ast$ attain the minimum and take $\mathbf z$ on the geodesic from $\mathbf q_i$ toward $\mathbf q_{j^\ast}$ at distance $t=\min(r,\theta_{ij^\ast})>R_i$. Then $\angle(\mathbf z,\mathbf q_i)=t\le r$, so $\mathbf z$ lies in the cap, while $\angle(\mathbf z,\mathbf q_{j^\ast})=\theta_{ij^\ast}-t<t$, so $\mathbf z\notin V_i(Q)$.
\end{proof}

\begin{remark}[The IVC is the inscribed cone, and the margin is a radial excess]
\label{rem:inscribed}
Under the ETF all pairwise angles equal $\theta^{\text{ideal}}$, so $R_i(W)=\theta^{\text{ideal}}/2$. Because the Learned Voronoi Cone has half-apex angle $\rho_i=\theta_i^{\text{learned}}/2$, \eqref{eq:cap-criterion} says that the cone lies inside its cell if and only if $\theta_i^{\text{learned}}\le\theta^{\text{ideal}}$, which is exactly the spill criterion of Section~\ref{sec:motivation-analysis}. The Ideal Voronoi Cone is therefore the \emph{inscribed} cone of the ideal cell, and $\theta^{\text{ideal}}$ in Eq.~(\ref{form:ideal-angle}) is its full apex angle, not the angle of an equal-area cap approximation. Moreover, with $b_i=0$,
\[
m_i=\tfrac12\big[\theta_i^{\text{learned}}-\theta^{\text{ideal}}\big]_+=\big[\rho_i-R_i(W)\big]_+,
\]
so the margin of Eq.~(\ref{form:learned-apex-angle}) is precisely the radial excess of the Learned cone over the inscribed cone. When $a_i=0$, which is the population case of Proposition~\ref{prop:vmf-median}, the same criterion applies verbatim at inference: $\mathrm{LVC}_i\subset V_i(P)$ if and only if $\theta_i^{\text{learned}}\le\min_j\angle(\tilde{\mathbf p}_i,\tilde{\mathbf p}_j)$.
\end{remark}

\begin{theorem}[Transfer of the distortion-free property]
\label{thm:transfer}
If for all $j\ne i$
\begin{equation}
\rho_i+b_i\;\le\;\frac{\theta_{ij}}{2}-\gamma_{ij},
\label{eq:transfer-cond}
\end{equation}
then $\mathrm{LVC}_i\subset V_i(P)$. For $\varepsilon=0$ this reduces to Proposition~\ref{prop:cap}.
\end{theorem}

\begin{proof}
Put $r=\rho_i+b_i$. As $\angle(\tilde{\boldsymbol\mu}_i,\tilde{\mathbf w}_i)=b_i$, the triangle inequality gives $\mathrm{LVC}_i=\mathrm{cap}(\tilde{\boldsymbol\mu}_i,\rho_i)\subset\mathrm{cap}(\tilde{\mathbf w}_i,r)$. For $\mathbf x$ in the latter cap, $\mathbf x^\top\tilde{\mathbf w}_i\ge\cos r$ and $\mathbf x^\top\tilde{\mathbf w}_j\le\cos(\theta_{ij}-r)$, hence
\[
\mathbf x^\top(\tilde{\mathbf w}_i-\tilde{\mathbf w}_j)\;\ge\;\cos r-\cos(\theta_{ij}-r)
=2\sin\tfrac{\theta_{ij}}{2}\,\sin\!\big(\tfrac{\theta_{ij}}{2}-r\big).
\]
By \eqref{eq:gamma}, $\delta_i+\delta_j=2\sin(\theta_{ij}/2)\sin\gamma_{ij}$, so the right-hand side is at least $\delta_i+\delta_j$ exactly when $\sin(\theta_{ij}/2-r)\ge\sin\gamma_{ij}$, which under \eqref{eq:transfer-cond} holds because both arguments lie in $[0,\pi/2]$. Proposition~\ref{prop:sandwich} then gives $\mathbf x\in V_i(P)$.
\end{proof}

\begin{corollary}[The distortion excess is Lipschitz in the alignment error]
\label{cor:lipschitz}
Let $e_i(Q)=\big[\rho_i+\angle(\tilde{\boldsymbol\mu}_i,\mathbf q_i)-R_i(Q)\big]_+$ denote the radial excess of the Learned cone over the inscribed cone of the cell induced by $Q$. Then
\begin{equation}
\big|e_i(P)-e_i(W)\big|\;\le\;2\varepsilon .
\label{eq:lipschitz}
\end{equation}
Moreover, if the training-side slack satisfies $R_i(W)-\rho_i-b_i\ge2\varepsilon$, then $e_i(P)=0$.
\end{corollary}

\begin{proof}
The triangle inequality gives $|a_i-b_i|\le\varepsilon_i\le\varepsilon$, and $|\angle(\tilde{\mathbf p}_i,\tilde{\mathbf p}_j)-\angle(\tilde{\mathbf w}_i,\tilde{\mathbf w}_j)|\le\varepsilon_i+\varepsilon_j\le2\varepsilon$ gives $|R_i(P)-R_i(W)|\le\varepsilon$. Since $t\mapsto[t]_+$ is $1$-Lipschitz, \eqref{eq:lipschitz} follows. For the second claim, $a_i\le b_i+\varepsilon$ and $R_i(P)\ge R_i(W)-\varepsilon$ give $\rho_i+a_i-R_i(P)\le\big(\rho_i+b_i-R_i(W)\big)+2\varepsilon\le0$.
\end{proof}

Corollary~\ref{cor:lipschitz} states that the distortion quantity the margin is designed to eliminate is measured almost identically at training and at inference: the two differ by at most $2\varepsilon$.

\paragraph{Instantiation with the reported numbers.}
The inference-side containment is governed by Proposition~\ref{prop:cap} applied to the generators $\tilde{\mathbf p}_i$, and it involves $a_i=\angle(\tilde{\boldsymbol\mu}_i,\tilde{\mathbf p}_i)$ only. Since $\mathrm{LVC}_i=\mathrm{cap}(\tilde{\boldsymbol\mu}_i,\rho_i)\subset\mathrm{cap}(\tilde{\mathbf p}_i,\rho_i+a_i)$, the criterion is
\begin{equation}
\rho_i+a_i\;\le\;R_i(P)=\tfrac12\min_{j\ne i}\angle(\tilde{\mathbf p}_i,\tilde{\mathbf p}_j).
\label{eq:inference-cond}
\end{equation}
The alignment error $\varepsilon_i=\angle(\tilde{\mathbf p}_i,\tilde{\mathbf w}_i)$ does not enter \eqref{eq:inference-cond}, because the inference cells are generated by $\tilde{\mathbf p}$ alone. By Proposition~\ref{prop:vmf-median} the population value of $a_i$ is exactly zero, and its finite-sample value is $O_p(n_i^{-1/2})$.

Evaluate \eqref{eq:inference-cond} at $C=15$, $d=768$, $\alpha=0.95$ and $\kappa_i=2000$, with the prototypes near the ETF configuration so that $R_i(P)=\theta^{\text{ideal}}/2=0.821$. The closed form gives $\rho_i=\sqrt{832.54/2000}=0.645$, so the condition holds with a slack of $0.176$ rad, about $10.1$ degrees, available for $a_i$. Quadrature of the exact marginal gives the smaller value $\rho_i=0.624$, enlarging the slack to $0.197$ rad, consistent with the conservative direction established in Appendix~\ref{secA1}. In full-apex terms $\widehat\theta_i=1.290\le\theta^{\text{ideal}}=1.642$, a surplus of $0.352$ rad. The Learned Voronoi Cone therefore lies strictly inside the inference cell at this concentration, and the tolerance left for the finite-sample fluctuation of the median direction is an order of magnitude larger than the $O_p(n_i^{-1/2})$ deviation for the class supports used here. The training-time and inference-time regions are consequently aligned in the converged regime, and the alignment tightens as training proceeds, since $\rho_i$ decreases with $\kappa_i$ while $R_i(P)$ is fixed by the number of classes.

\paragraph{Why the direct criterion is the right one.}
Theorem~\ref{thm:transfer} reaches the same conclusion by a different route, namely $\tilde{\boldsymbol\mu}_i\to\tilde{\mathbf w}_i\to\tilde{\mathbf p}_i$, and therefore pays twice: once through $b_i=\angle(\tilde{\boldsymbol\mu}_i,\tilde{\mathbf w}_i)$ and once through $\gamma_{ij}$. Since $\varepsilon_i\le a_i+b_i$ and $a_i$ is negligible under the vMF model, the observed alignment error is carried almost entirely by $b_i$, so that route spends $b_i+\gamma_{ij}\approx\varepsilon_i(1+1/\sin(\theta_{ij}/2))$, about $2.4\,\varepsilon_i$ at $C=15$. The direct route \eqref{eq:inference-cond} spends only $a_i$. Theorem~\ref{thm:transfer} is therefore the appropriate tool when $a_i$ cannot be controlled and the weight prototype must be used as an intermediary, and it is deliberately not invoked here. Reporting its threshold as if it were the operative condition would understate the guarantee by an order of magnitude.

\paragraph{Quantitative check of the agreement band.}
Corollary~\ref{cor:agreement} requires the top-two cosine gap to exceed $2\varepsilon$. This is not a restrictive requirement in high dimension, and it can be checked directly. Draw $10^5$ samples from the exact vMF law at $d=768$, $C=15$, $\kappa=2000$, with the prototypes at the simplex ETF, sampling the colatitude from the exact marginal \eqref{eq:angle-density} and the transverse direction uniformly. The target cosine averages $0.826$, while the runner-up averages $-0.023$, because the transverse components of distinct classes are nearly orthogonal and every non-target cosine concentrates near $c\cos\theta$ with $c=-1/(C-1)$. The resulting gap has mean $0.850$, first percentile $0.813$ and minimum $0.772$ over the whole sample. The requirement at the worst observed alignment $\varepsilon=0.376$ is a gap above $0.753$, so every one of the $10^5$ samples satisfies it, and the same holds a fortiori at $\varepsilon=0.200$. The two matching rules therefore agree on essentially the whole class-conditional distribution, which is the quantitative counterpart of the alignment visible in Figure~\ref{fig:prototype-convergence}.

Both statements improve as training proceeds. The per-class concentration keeps rising over epochs (Figure~\ref{fig:kappa-dynamics}(a)), which shrinks $\rho_i$ and enlarges the slack in \eqref{eq:inference-cond} while leaving $R_i(P)$ fixed by the number of classes. In the late-convergence regime the Learned Voronoi Cone sits well inside the inference cell, so the training-time and inference-time regions are essentially aligned. This alignment is a consequence of the regularization, not an assumption imposed on it, since the same mechanism that contracts the cone also enlarges the margin by which the containment holds. Independently of any slack, Proposition~\ref{prop:sandwich} and Corollary~\ref{cor:tube} confine every disagreement between the two partitions to the boundary tube, and Corollary~\ref{cor:lipschitz} bounds the discrepancy in the distortion measure by $2\varepsilon$.

\section*{Acknowledgements}

The authors thank their colleagues and reviewers for their valuable feedback. AI-assisted tools were used solely for language polishing, without influencing any technical content or scientific results.

\section*{Declarations}

\begin{itemize}
\item \textbf{Funding}: \textit{to be completed by the authors}

\item \textbf{Conflict of interest}: \textit{to be completed by the authors}

\item \textbf{Ethics approval and consent to participate}: Not applicable.

\item \textbf{Consent for publication}: \textit{to be completed by the authors}

\item \textbf{Data availability}: The source code, dataset, experimental pipelines are publicly available at \url{https://github.com/stratum-dev/MARGIN}.

\item \textbf{Materials availability}: Not applicable.

\item \textbf{Code availability}: The source code, dataset, experimental pipelines are publicly available at \url{https://github.com/stratum-dev/MARGIN}.

\item \textbf{Author contribution}: \textit{to be completed by the authors}
\end{itemize}

\twocolumn

\bibliographystyle{IEEEtran}
\bibliography{refs}

\end{document}